%% file: main.tex
\documentclass[fleqn,10pt]{wlscirep}
\usepackage[utf8]{inputenc}
\usepackage[T1]{fontenc}

\input{acronyms}

\input{math-definitions}
\usepackage{orcidlink} 

\title{System Characterization of a Human-Sized 3D Real-Time Magnetic Particle Imaging Scanner for Cerebral Applications}

\author[1,2,*,+]{Florian Thieben~\orcidlink{0000-0002-2890-5288}} 
\author[1,2,+]{Fynn Foerger~\orcidlink{0000-0002-3865-4603}} 
\author[1,2,+]{Fabian Mohn~\orcidlink{0000-0002-9151-9929}}
\author[1,2]{Niklas Hackelberg~\orcidlink{0000-0002-0976-9049}} 
\author[1,2]{Marija Boberg~\orcidlink{0000-0003-3419-7481}} 
\author[3,4]{Jan-Philipp Scheel~\orcidlink{0000-0002-7840-4879}} 
\author[1,2]{Martin Möddel~\orcidlink{0000-0002-4737-7863}}
\author[1,2,3,4]{Matthias Graeser~\orcidlink{0000-0003-1472-5988}} 
\author[1,2,3]{Tobias Knopp~\orcidlink{0000-0002-1589-8517}} 

\affil[1]{Section for Biomedical Imaging, University Medical Center Hamburg-Eppendorf, Hamburg, Germany}
\affil[2]{Institute for Biomedical Imaging, Hamburg University of Technology, Hamburg, Germany}
\affil[3]{Fraunhofer Research Institute for Individualized and Cell-based Medicine, Lübeck, IMTE,  Germany}
\affil[4]{Institute for Medical Engineering, University of Lübeck, Lübeck, Germany}

\affil[*]{f.thieben@uke.de}

\affil[+]{these authors contributed equally to this work}


\begin{abstract}
Since the initial patent in 2001, the \ac{MPI} community has been striving to develop an \ac{MPI} scanner suitable for human applications. Numerous contributions from different research fields, regarding tracer development, reconstruction methods, hardware engineering, and sequence design have been employed in pursuit of this objective.
In this work, we introduce and thoroughly characterize an improved head-sized \ac{MPI} scanner with an emphasis on human safety. The scanner is operated by open-source software that enables scanning, monitoring, analysis, and reconstruction, designed to be handled by end users. 
Our primary focus is to present all technical components of the scanner, with the ultimate objective to investigate brain perfusion imaging in phantom experiments. 
We have successfully achieved full 3D single- and multi-contrast imaging capabilities at a frame rate of 4\,Hz with sufficient sensitivity and resolution for brain applications. 
To assess system characterization, we devised sensitivity, resolution, perfusion, and multi-contrast experiments, as well as field measurements and sequence analysis. 
The acquired images were captured using a clinically approved tracer and suitable magnetic field strengths, while adhering to the established human \acl{PNS} thresholds. 
This advanced scanner holds potential as a tomographic imager for diagnosing conditions such as ischemic stroke or intracranial hemorrhage in environments lacking electromagnetic shielding. 
Furthermore, due to its low power consumption it may have the potential to facilitate long-term monitoring within \aclp{ICU} for various applications. 
\end{abstract}
\begin{document}

\flushbottom
\maketitle

\thispagestyle{empty}

\section{Introduction}
\acresetall 
Human-scale \acf{MPI} has the potential to provide background- and radiation-free tomographic images, at high temporal resolution.
In comparison, \ac{CT} provides high resolution images at the cost of radiation exposure and is ill-suited for reoccurring or long-term monitoring, whereas \ac{MRI} suffers from limited patient accessibility and lengthy scan durations.
Both modalities come as large and fixed systems, which makes it challenging to use them in an \ac{ICU}. The main challenge is that patients usually need to be transported, which is not without risks and requires considerable preparation~\cite{porte_imaging_2009}. In addition, patients are highly encapsulated in MRI systems, making it difficult to monitor their condition~\cite{rotello_mri_1994}. These are also the reasons why portable devices such as chest radiograph and ultrasound are mainstay imaging modalities in \acp{ICU}~\cite{lohan_imaging_2019}.
MPI also has great potential in this direction, due to its size and compatibility with the \ac{ICU} environment. 
\ac{MPI} brain images could be acquired inside the \ac{ICU}, directly at the patient's bed, reducing the workload for medical staff, avoiding transport of the patient and shortening time of treatment decisions. 
As a quantitative and tracer-based imaging modality, \ac{MPI} is able to visualize the structure of larger blood vessels or quantify tissue perfusion with high temporal resolution, among other diagnostic and therapeutic applications~\cite{ludewig_magnetic_2022}.
The potential application in \acp{ICU} targets neurovascular diseases like ischemic stroke, intracranial hemorrhage and traumatic brain injury that require immediate attention and post-treatment monitoring to evaluate the procedure. More than 17 million cases occur each year worldwide and are a leading cause of death and disability~\cite{feigin_global_2014,graeser_human-sized_2019}, motivating further research and development.

Currently, \ac{MPI} is in the process of upscaling~\cite{knopp_magnetic_2017} the pre-clinical (small rodent) bore size~\cite{weizenecker_three-dimensional_2009,vogel_first_2016} to match human proportions like the head~\cite{borgert_perspectives_2013,graeser_human-sized_2019,mason_design_2017} or extremities~\cite{vogel_impi_2023}. Images are acquired with high spatiotemporal resolution, providing background-free contrast information, based on the non-linear response of \ac{MNPs}~\cite{gleich_tomographic_2005} with high sensitivity~\cite{graeser_towards_2017,graeser_design_2020}. 
Various combined magnetic fields in the low \si{\kHz} range allow the spatial selection and detection of a tracer, such as the established contrast agent Ferucarbotran from \ac{MRI}, with up to 46 volumes per second~\cite{weizenecker_three-dimensional_2009}.
\ac{MPI} was shown to be capable of hyperthermia treatment~\cite{murase_control_2013,tay_magnetic_2018}, stem cell labeling~\cite{bulte_quantitative_2015}, detection of lymph node metastatsis~\cite{harisinghani_noninvasive_2003}, gut bleeding~\cite{yu_magnetic_2017} and lung perfusion imaging~\cite{ludewig_magnetic_2017} in murine models, cancer detection~\cite{yu_magnetic_2017-1}, as well as being useful in interventional applications that involve guiding catheters~\cite{rahmer_interactive_2017,herz_magnetic_2018} and stent positioning~\cite{ahlborg_first_2022}.
In addition, the \ac{MPI} tracer can be used as a micro probe for several external and internal parameters that change the signal response, like the carrier medium viscosity~\cite{moddel_viscosity_2018} and temperature~\cite{stehning_simultaneous_2016}. Furthermore, particle properties like binding state~\cite{viereck_dual-frequency_2017}, the particle core-size~\cite{rahmer_first_2015, shasha_discriminating_2019} or the orientation~\cite{moddel_estimating_2020}, can be derived and visualized using multi-contrast imaging.

One objective of upscaling \ac{MPI} scanners is to investigate clinical utility by evaluating the power required to generate the magnetic fields and by assessing realistic resolution and sensitivity of real-time imaging at this scale. A major challenge is to minimize the scanners power consumption, especially of the \ac{SF}~\cite{foerger_flexible_2023}, as well as ensuring patient safety in proximity to high power components. Our approach encompasses these conditions and attempts to combine standard-socket power supply, an unshielded environment and to meet local medical device safety regulations.

In this paper, we describe a human-sized MPI system for brain applications, and verify its functionality in several experiments using a clinically approved tracer~\cite{hartung_resotran_2023}.
The general scanner concept is based on Graeser et al~\cite{graeser_human-sized_2019}, although many hardware components were replaced, improved or re-developed for this version to increase instrumental safety with a focus on future human trials. A major improvement is the realization of 3D imaging by using 2D excitation from two orthogonal \ac{DFCs} and a slow shift of the dynamic selection field to achieve volumetric sampling at \SI{4}{\Hz}.
In this work, we describe and analyze the full system design and implementation including our excitation fields and the measured system matrix, as well as characterize the overall performance of the brain scanner with resolution-, sensitivity-, perfusion-, and multi-contrast-experiments.

Our work represents a significant step towards clinical application of \ac{MPI} and may pave the way for monitoring neurovascular diseases within the \ac{ICU}. The scanner system described is suited for end users, with an adaptable and interactive \ac{GUI}, an open source reconstruction framework and redundant safety mechanisms, that facilitate performing \ac{MPI} scans and the live inspection of results.
We elaborate on all system components, explain how the magnetic fields are generated, provide insight to our custom signal receive chain, and characterize the imaging performance using the system matrix approach.

\renewcommand{\thefootnote}{\textit{\alph{footnote}}}

\section{System Overview}
\label{sec:overview}

\begin{figure}[t!]
    \centering
    \includegraphics[width=1.0\textwidth]{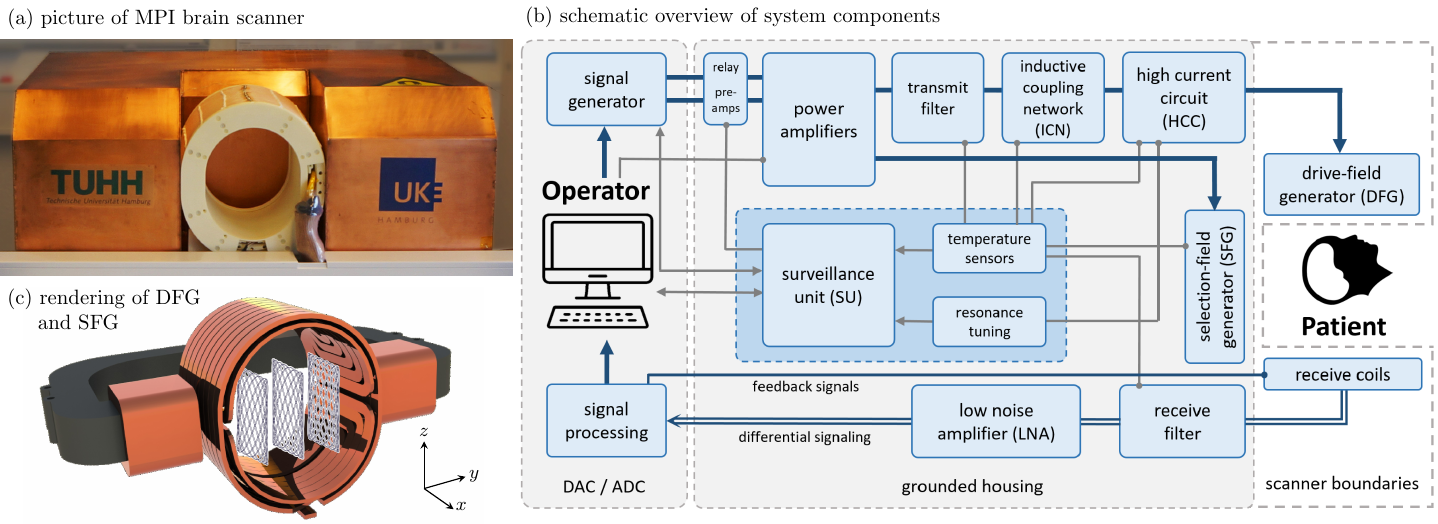}
    \caption{A photo of the MPI brain scanner is shown in (a), with the \acf{DFG} in a white 3D printed housing. In (b), an schematic overview of the entire scanner is provided by a flowchart showing the key system components and their interactions.
    A rendering of the \ac{DFG} is shown below in (c), including the \ac{SFG}, which is an iron yoke that creates the dynamic selection field. Individual closed (ideal) 2D Lissajous trajectories are visualized in the $xz$-plane, that are shifted by \ac{FFP} displacement in $y$-direction in the selection field.}
    \label{fig:Overview}
\end{figure}

The presented MPI brain scanner is captured in an image in \autoref{fig:Overview}\,(a) and a schematic block diagram is depicted in (b). The scanner can be classified into four main parts: 
operational control (\autoref{sec:field:OCU}), 
field generation (\autoref{sec:field:DF}, \ref{sec:field:selec}, \ref{sec:fields:Seq}), 
signal reception (\autoref{sec:field:rx}), 
and data processing (\autoref{sec:data}), which are briefly introduced in this section for an overview. 

\paragraph{The operational control} is tasked with coordinating signal generation (\ac{DAC}) and reception (\ac{ADC}). The system devices are driven and coordinated by a collection of open-source software~\cite{hackelberg_flexible_2022,hackelberg_mpimeasurementsjl_2023}. The combined software-stack realizes a framework, which is adaptable and scalable to many different MPI applications.
The framework accepts user inputs via a command line interface, and more conveniently, via a \ac{GUI}\footnote{https://github.com/MagneticParticleImaging/MPIUI.jl}, that allows different types of measurements to be started, paused, aborted, or stopped. Also, it enables live displaying and analyzing measurement data and controlling and monitoring system devices such as temperature sensors or robots.
The acquired MPI signals are stored in the open MPI data format (MDF)~\cite{knopp_mdf_2016}.
Another part of the operation control is a \ac{SU} for monitoring, based on the micro-controller board \textit{Arduino Mega 2560 Rev3}. It processes sensor data, controls the state of signal relays, and communicates via analog pins with amplifiers, temperature units, and control units, as well as serial communication with the operational software. Some functions, status information and relays are routed to a hardware user console, placed with the human operator, allowing interaction via buttons to activate or deactivate key parts of the system at will.

\paragraph{Field generation} is initiated by four \acp{DAC}, which generate two \ac{DF} signals and the two signals for the dynamic selection field.
The \ac{DF} signal of each channel is connected to a floating transmission chain with a $5$th-order band-pass filter that powers the \ac{DFG} via an \ac{ICN}.
The \ac{DFG} is responsible for 2D field excitation in the $xz$-plane (\SI{5}{\milli\tesla}, \SI{4}{\milli\tesla})
and forms a high quality-factor resonator that is designed to carry high currents at low voltage (referred to as \ac{HCC}).
The \ac{ICN} is a toroidal transformer that serves three purposes, high current gain, circuit symmetry, and floating potentials to increase human safety.
The dynamic selection field is generated by the \ac{SFG} with two coils mounted on an iron yoke inside a copper cabin, identical to this part of the setup in Graeser et al~\cite{graeser_human-sized_2019}. Identical \SI{10}{A} coil currents generate a \SI{0.24}{\tesla\per\meter} gradient field with a \ac{FFP} in the center of the \ac{FOV}.
By varying these currents, the \ac{FFP} is moved along the $y$-direction to create a large nominal 3D \ac{FOV} (\qtyproduct{84x85x67}{\milli\meter}). The dynamic selection-field current waveform can be chosen to be sinusoidal, or in our case, triangular, for a constant shifting motion with a \SI{4}{\Hz} imaging sequence. This principle is shown in \autoref{fig:Overview}\,(c) and explained along with all mentioned fields in more detail in \autoref{sec:field}.

\paragraph{Signal reception} is realized with a gradiometric receive coil for feedthrough suppression in $x$-direction and a saddle coil in $y$-direction. Given that both excitation coils are orthogonal to the $y$-direction, this channel is not required to suppress high feedthrough signals by gradiometric turns. Both receive coils are connected to a symmetric $4$-th order band-stop filter, transferred to single ended signals via a balun transformer matching a single ended custom \ac{LNA}. 
Signals are then transferred differentially towards the \acp{ADC}, which are configured with a receive bandwidth of \SI{976.56}{\kHz}. Furthermore, two more signals are passed to the \acp{ADC} for feedback during the control-step at the beginning of measurements, that automatically scales the transmit channel currents to match the desired \ac{DF} strengths.

\paragraph{Data processing} describes image reconstruction and further processing steps in the case of perfusion images. For image reconstruction, the system matrix approach is used, implemented by the open-source MPI reconstruction framework~\cite{knopp_mpirecojl_2019}. The measured data can be interpreted as multi-patch data corresponding to the selection-field shifts, or as a 3D single-patch dataset. We use a Kaczmarz-solver, system matrix over-gridding (interpolation), $L^2$-as well as $L^1$-regularization, background subtraction, and frequency selection to obtain images (see \autoref{sec:data:recon} for details). These are presented for different experiments with individually tuned reconstruction parameters in  \autoref{sec:results}. Perfusion images are calculated based on reconstructed 3D volumes that are filtered and processed to obtain the \ac{TTP}, \ac{MTT}, \ac{rCBF} and \ac{rCBV}. The definition and details on the implementation can be found in \autoref{sec:data:perfcalc}.

\section{Field Generation and Reception}
\label{sec:field}

This section provides a detailed presentation of all scanner components and introduces the concept of how they interact to obtain \ac{MPI} measurements. Beginning with the generation of magnetic fields, details are given on signal reception, imaging sequences, operation control, and data processing for image reconstruction and post-processing.

The scanners fundamental component for signal generation and acquisition is realized by a stack of three \ac{RPs}. These are a flexible low-cost hardware solution for integrating \acp{DAC} and \acp{ADC} into a single device for precise real-time signal handling. The open-source software for the \ac{RPs} by Hackelberg et al.~\cite{hackelberg_flexible_2022}, ensures a parallel and synchronous signal generation and reception by synchronizing the \SI{125}{MHz} clock and logic of the three used \ac{RPs}. This stack thus provides six radio-frequency input and output channels.

\subsection{Drive-Field Generation}
\label{sec:field:DF}

\begin{figure}[!t]
    \centering
    \includegraphics[width=1.0\textwidth]{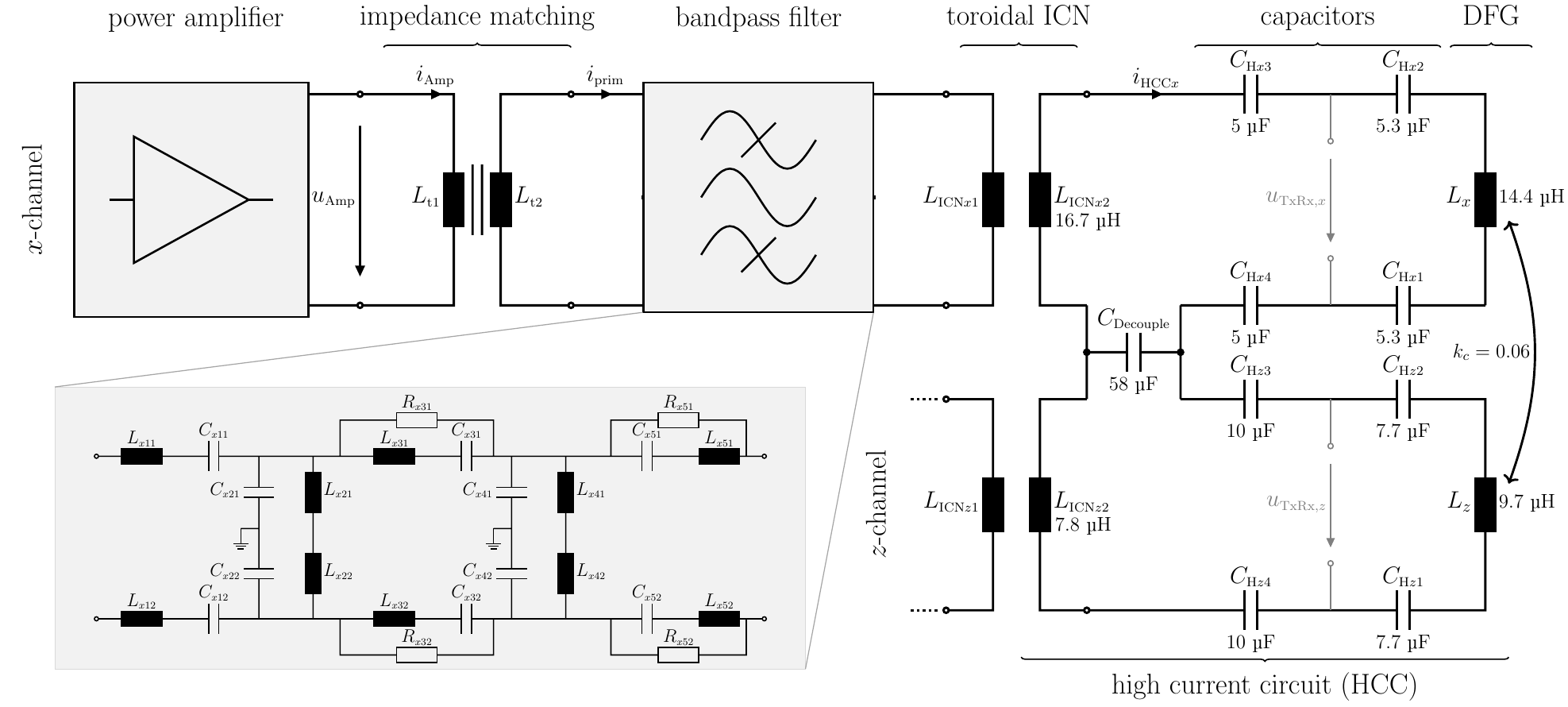}
    \caption{Simplified transmit circuit for $x$- and $z$-\ac{DF} excitation at \SI{25.699}{\kHz} and \SI{26.042}{\kHz}, respectively. Both channels consist of their own power amplifier, impedance matching transformer (with iron core, ||) and bandpass filter. The bandpass filter is shown for $x$ only, but is identical for the $z$-channel. The \ac{ICN}s were individually developed with different parallelizations (higher current in $z$) and roughly match the inductance of the corresponding DFC. The entire \ac{HCC} is resonant. Residual coupling of the \ac{DF} channels and decoupling counter-measures are also shown. Values of major components are denoted in the \acl{ECD}. Since fine tuning needs to be performed with a fully assembled \ac{HCC}, the denoted \ac{HCC} capacity values are guide values.}
    \label{fig:Tx}
\end{figure}

In \autoref{fig:Tx}, a simplified \acf{ECD} of the transmit circuit is shown, with details of the transmit filter for one of the two channels. In the following, we describe the components of the transmit chain in detail from field to source: the \acf{DFG}, the \acf{HCC}, the \acf{ICN}, the transmit filter and the impedance matching transformer at the output of the amplifier.

\paragraph{Drive-field generator:} The MPI brain scanner utilizes two \ac{DFCs} in $x$- and $z$-direction, respectively. 
The solenoid $x$-DFC (with inductance $L_x$) and the saddle coil $z$-DFC (with inductance $L_z$) are nested, with the $x$-DFC placed on the inside. Both are manufactured using Rutherford wire parallelization with 12 individual strands of a high frequency Litz wire (2000 isolated strands with \SI{50}{\micro m} diameter, Elektrisola, Germany). It further serves the minimization of the self-inductance of the wire, keeps individual wires at identical lengths, and mitigates proximity- and skin-effect. By parallelizing 12 Litz wires, a minimized serial resistance is achieved while maintaining the ability to form and wind the wire into the desired coil topology~\cite{ozaslan_design_2020}. In the \ac{FOV} center, the manufactured \ac{DFCs} exhibit coil sensitivities of \SI{0.022}{\mT\per\A} and \SI{0.014}{\mT\per\A} in $x$- and $z$- direction, respectively. Further details on the achieved component values can be found in \autoref{fig:Tx}.
The support structure (PA2200) of the \ac{DFG} forms an elliptic open bore with a width of \SI{17.5}{cm} and a height of \SI{21.5}{cm}.
The resulting 2D \ac{DF} is set up in the $xz$-plane with \ac{DF} frequencies set to $f_x = \frac{125}{4864}~\si{\MHz}\approx\SI{25.699}{\kHz}$ and $f_z=\frac{125}{4800}~\si{\MHz}\approx\SI{26.042}{\kHz}$. 
The frequency ratio between the two channels results in a closed 2D Lissajous trajectory after $76$ and $75$ periods for the $x$- and $z$-directions, respectively.

\paragraph{High current circuit:} In order to mitigate high voltages near the human brain, a design objective was to utilize parallelized low inductance \ac{DFCs} that require high currents instead of high voltages for generating the desired drive field. To minimize reactive power and obtain symmetry, the \ac{DFCs} are operated in resonance at the \ac{DF} frequency by connecting two capacitors of equal size both upstream and downstream of the inductance. Instead of utilizing a capacitive voltage divider for impedance matching~\cite{mattingly_drive_2022}, a mirrored resonant setup with a toroidal transformer coil ($L_{\textup{ICN}2}$) is connected to each DFC with the same resonance frequency. For resonance tuning, the capacities $C_{\textup{H}x_i}, C_{\textup{H}z_i}, i=1,...,4$ are temperature controlled  capacitors (CSM 150/200, Celem, Israel), which enables stable and precise resonance tuning~\cite{thieben_heat_2022}. 

Manufacturing orthogonal \ac{DFCs} is challenging, hence residual coil coupling between \ac{DF} channels must be addressed in order to avoid beat frequencies, undesired frequency shifts by mode splitting, and resulting losses.
Due to the resonant behavior of the coupled coil circuit and the low difference of \ac{DF} frequencies, the coupled signals also experience an amplification in the other circuit. 
Even the small coupling coefficient of $k_c = 0.06$ can lead to large currents in the orthogonal coil, resulting in a strong beating signal of the two frequencies. For the \ac{HCC}, the field coupling far exceeds \SI{10}{\percent}, causing a distorted trajectory, which significantly reduces the image quality~\cite{von_gladiss_influence_2018}. Coupling is reduced by a decoupling capacitor ($C_\textup{Decouple}$) that is connected in series within both circuits. It carries both currents and matches the complex conjugate impedance of the mutual inductance $L_\textup{M} =  k_c\,\sqrt{L_x~ L_z}$ between the two \ac{DFCs}. 

As a result, the voltage drop across the capacitor cancels the induced voltage from the other channel within each channel loop. For our setup, we built a capacitor $C_\textup{Decouple} \approx \SI{58}{\micro\farad}$ by parallelization of $58\cdot \SI{1}{\micro\farad}$ (MKP C4Q, Kemet, USA) and achieve a residual channel coupling of \SI{-35}{\dB}.

\paragraph{Inductive coupling network:} 
The aforementioned toroidal air-core transformer coil composes the \acf{ICN}, one for each \ac{DF} channel. The \ac{ICN} serve the triple purpose of high current gain (impedance matching), circuit symmetry, and floating potentials. The last point is achieved by any transformer. It increases patient safety by requiring direct contact to two separate points of the circuit in order to cause a voltage drop across a grounded person. Also, a floating \ac{HCC} avoids ground-loops that may have a negative influence on signal reception. The intention of circuit symmetry is to obtain two voltage nodes in each channel, between which the voltage of the fundamental frequency is zero. Both inductors of a channel (\ac{ICN} and \ac{DFG}) are in a series resonance with their corresponding capacitors, which have equal voltages but opposing sign at resonance. A pick-up point for the particle harmonics is thus created, which is nested in-between the capacitors within the \ac{HCC}, denoted by $u_\textup{TxRx}$ in \autoref{fig:Tx}, with a suppressed feedthrough~\cite{sattel_setup_2015}.
For the higher harmonics induced by the particles, the inductors on both sides result in higher impedances, creating an inductive voltage divider for both noise and particle signals. The \ac{ICN} also plays a crucial role in providing current gain at the resonance of the \ac{HCC}. This enables the transformation of filtered transmit signals into the desired low-voltage-high-current signal. 
Multiple Litz-wire parallelization on the secondary transformer side ($L_\textup{ICN2}$) are used to achieve low losses and a high quality factor, nevertheless, additional inductance and resistance is introduced by the \ac{ICN} in the resonant \ac{HCC}.
The \ac{HCC} constitutes the load of the transformer, and the primary transformer coil ($L_\textup{ICN1}$) shows a real load impedance ($Z_\textup{ICN} \approx \SI{30}{\ohm}$) at resonance frequency and becomes part of the last stage of the transmit filter. The achieved current gain is $31$ for the $x$-channel and $35$ for the $z$-channel.

\paragraph{Transmit filter:} A differential $5$th-order band-pass filter is used to smoothen the excitation voltage before it is connected to the primary side of the \ac{ICN}, shown in \autoref{fig:Tx}. The three odd stages of the filter consist of serial resonance circuits that act as a band-pass by having a minimum impedance at resonance. In-between these stages, two even stages act as parallel resonators that create a short for all other frequencies in order to dissipate the energy of undesired frequencies. The serial resonator (odd stages) is composed of a 3D printed toroidal air coil ($L_\text{s,odd} \approx $ \SI{500}{\micro\henry}, $R_\text{s,odd} \approx$ \SI{1}{\ohm}). For a symmetrical differential signal, the toroid is separated into two halves, with each half forming their own resonance together with a high voltage, polypropylene film capacitor (e.g. KEMET C4C series, 2 kV DC rated). Due to the mutual magnetic field within the toroidal coil, the field lines run through both sides of the toroid. Additional resistors across the second and third serial resonator stages are utilized to attenuate side lopes~\cite{mattingly_drive_2022}.
The two parallel resonators (even stages) are formed by smaller toroidal air coils ($L_\text{s,even} \approx $ \SI{250}{\micro\henry}, $R_\text{s,even} \approx$ \SI{500}{\milli\ohm}) to pose a high impedance at the \ac{DF} frequency between the two voltage rails. The assembled transmit-filter chains attenuate harmonic distortions of the \ac{DF} signal by $-65$\,dB, $-100$\,dB, and $-150$\,dB amplitude ratio for the second, third, and fourth harmonics, respectively. The overall differential filter setup provides common mode rejection.

\paragraph{Transformer:} For maximum power transfer, we implemented an impedance matching transformer with an iron core (N87 material B65686A0000R087, TDK Electronics, Germany).
To prevent distortion and harmonics due to core saturation effects, the core flux density is minimized to \SI{16}{\%} of the saturation magnetization $B_\text{sat}$\cite{mattingly_drive_2022}.
The secondary side of the transformer is connected to the transmit filter and ensures floating potentials and differential signaling. The turns ratio of the transformer changes the low impedance ($\approx$ \SI{2.5}{\ohm}, amplifier side) to a high impedance ($\approx $\SI{30}{\ohm}, filter side) to minimize the current in the transmit filter.

\paragraph{Drive-field power amplifier:} Two \SI{1200}{\W} power amplifier (A1110-40-QE-100, Dr. Hubert GmbH,  Germany) are used in voltage mode for amplification of each \ac{DF} signal. To generate the \ac{DF} strength of \SI{5}{\milli\tesla} and \SI{4}{\milli\tesla}, a total power of \SI{930}{W} for $x$ and \SI{1100}{W} for $z$ has to be provided by the power amplifiers.
For safety and control reasons, the initial signal from the \ac{DAC} runs through a relay at the input of the amplifier that is only closed during measurements (by the \ac{SU}). In addition, serial interlock commands are sent to the amplifier to ensure that the output is only enabled during a measurement. The advantage of this double safety configuration is that the \ac{SU} is able to interrupt transmission in case of failure (e.g. temperature overshoot), even when the software activated the amplifier via the interlock mechanism. A pre-amplifier is used to amplify the voltage signal by a factor of 8 to scale it to the required input voltage of the power amplifier. 

\paragraph{Feedback signal:} To control the \ac{DF} signal in amplitude and phase, one turn is wound around each \ac{ICN} toroid. According to the law of induction, the induced voltage is proportional to the field, which is proportional to the \ac{HCC} current and hence to the drive field. The induced signal is fed back to the \ac{RPs} via a voltage divider, where it is processed during the \ac{DF} control phase. To generate a stable \ac{DF} trajectory, the control accuracy is set \SI{<1}{\%} and able to generate field strengths up to \SI{5.5}{\mT} for the $x$-channel and \SI{4.5}{\mT} for the $z$-channel.

\subsection{Selection-Field Generation}
\label{sec:field:selec}
The previously presented \ac{SFG} in Graeser et al.~\cite{graeser_human-sized_2019} is used to generate the required gradient field and further moves the \ac{FFP} in $y$-direction with a maximum displacement amplitude of \SI{70}{\mm} from the center.
The \ac{SFG} consists of two coils ($L_\textup{G} = \SI{200}{\milli\henry}$) mounted on a soft iron yoke at a distance of \SI{31}{cm} from each other. The magnetic gradient field is generated by superimposing the two field contributions, in a Maxwell-like coil topology with opposing current directions. Setting of equal opposite coil currents creates an \ac{FFP} in the \ac{FOV} center.
By changing the currents, the position of the \ac{FFP} along the $y$-axis can be shifted and the maximum \ac{FFP} offset is set to $\SI{43}{mm}$ to each side with respect to the \ac{FOV} center.
We chose the \ac{FFP} velocity to be $\SI{68.8}{\cm\per\s}$, which results in a similar trajectory density in $y$-direction as in the $xz$-plane (see \autoref{sec:res:SManalysis}).
The selection-field signals are generated by the aforementioned \acf{RPs}. A pre-amplifier is used to scale the input voltage according to the required voltage for the power amplifier for selection-field generation. For this purpose, two AE Techron 2105 (AE Techron, USA) in current mode feed the selection-field coils. For the generation of the gradient, a total power of \SI{380}{W} is required.

\begin{figure}[b!]
    \centering
    \includegraphics[width=\textwidth]{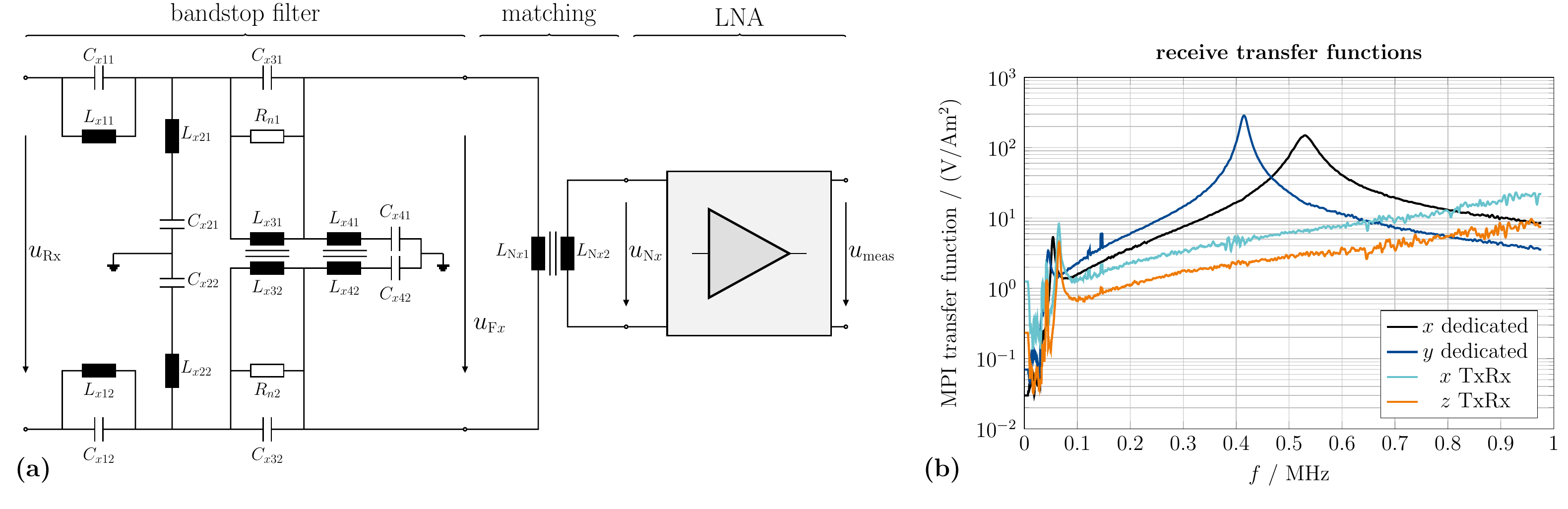}
    \caption{\ac{ECD} of a receive filter circuit with signal matching for the \acf{LNA}. In the receive circuit, a 4-stage analog bandstop filter is installed for $x$-, $y$-, and $z$-signal reception, although only $x$ and $y$ are used in this work. The first two filter stages are utilized by air coil resonators and the last two stages by iron core resonators. A signal-matching transformer follows the filter to adjust the receive signal prior to the \ac{LNA}. For differential signaling, the \ac{LNA} output is connected to a differential amplifier and in front of the \ac{ADC}, an instrumentation amplifier generates the single ended measurement signal $u_\text{meas}$. 
    On the right in (b), the receive path MPI transfer function for the transformation of $u_\text{meas}$ into the magnetic moment $m_\text{meas}$ is plotted for the two dedicated receive coils and the two send-receive approaches (TxRx).}
    \label{fig:RxFilter}
\end{figure} 

\subsection{Signal Reception}
\label{sec:field:rx}

\paragraph{Receive coils:} 
Two dedicated receive coils are used for signal reception for all measured data in this work.
The gradiometric $x$-coil has \num{10} turns on a length of \SI{10}{\cm} and \num{18} counter turns at the rear end of the \ac{DFG}.
In $y$-direction, a saddle coil with two times \num{20} turns is installed. Due to the orthogonal orientation to the two \ac{DFCs}, a gradiometric setup is not required and residual feedthrough voltages are suppressed by the receive filter. Both dedicated receive coils are located within the 3D printed housing of the \ac{DFG}, on the inside of the \ac{DFCs}.
A dedicated $z$-receive coil has not yet been implemented, due to its challenging design as a gradiometric saddle coil. The small homogeneous field region of the $z$-channel does not allow a clear spatial separation of the \ac{FOV} and the sensitive region of the gradiometer. 

However, as mentioned in \autoref{sec:field:DF} in the \ac{ICN} paragraph, the symmetric circuit design included a second method for signal pick-up, referred to as the send-receive method (TxRx). Consequently, receive signals for $x$ and $z$ can be obtained in principal. The pick-up node is positioned in the LC circuits of the \ac{HCC} between $C_\text{H2}$, $C_\text{H3}$ and $C_\text{H1}$, $C_\text{H4}$, for each \ac{DF} channel (see \autoref{fig:Tx}). In theory, the voltage of the fundamental excitation frequency is near zero at this node and any induced particle signal will create a voltage drop across the DFC (and the parallel \ac{ICN}, halving the receive signal if $L_{\textup{ICN}x2}\approx L_x$, or $L_{\textup{ICN}z2}\approx L_z$). Between these nodes, the input of the receive-chain can be connected, enabling signal reception during transmission~\cite{sattel_setup_2015}.

\paragraph{Receive chain:}
The receive voltage $u_\textup{Rx}$ is composed of the particle signal, the background noise, distortions and the direct signal feedthrough from the \ac{DFCs}. Before the signal is connected to the \ac{LNA}, a $4$th-order resonance notch filter suppresses the fundamental frequencies around \SI{26}{\kHz} (\SIrange{20}{33}{\kHz} stopband). The filter is a differential circuit, shown in \autoref{fig:RxFilter}\,(a). The odd stages feature a parallel resonance with high impedance at the \ac{DF} frequency and the even stages are serial resonances towards ground with low impedance at the \ac{DF} frequency. To avoid nonlinear effects due to high receive voltages, the first two stages utilize toroidal air coil resonators. 
The attenuated residual voltage in the last two stages allows the use of ferrite iron core coils (B64290, TDK Electronics, Germany).
The filtered receive signal is connected to an improved version of the custom \ac{LNA} by Graeser et al.~\cite{graeser_human-sized_2019} via a signal-matching transformer. The transformer and the input impedance of the \ac{LNA} are used to shift the resonance frequency of \ac{LNA} and receive coil to optimize signal amplification of higher harmonics in the \SIrange{300}{600}{\kHz} range~\cite{zheng_optimal_2017}. The \ac{LNA} consists of three amplification stages, the first stage is built by parallelizing $20$ JFETs (BF862, NXP Semiconductors, Netherlands), followed by two stages with non-inverting operational amplifier circuit using dual ICs (ADA4898, Analog Devices, USA)~\cite{mattingly_sensitive_2022}. The output is converted into a differential signal by a fully differential amplifier (AD8138, Analog Devices, USA). Just before the input of the \ac{ADC} of the \ac{RPs}, an instrumentation amplifier (AD8253, Analog Devices, USA) generates the single ended measurement signal $u_\textup{meas}$.

For each of the four receive chains, the receive path calibration was performed utilizing a custom calibration coil~\cite{thieben_receive_2023}. In \autoref{fig:RxFilter}\,(b) the individually measured MPI transfer functions are plotted. 
The two dedicated receive coils provide resonances at \SI{531}{\kHz} and \SI{414}{\kHz} for $x$ and $y$ respectively. It can be seen that the TxRx signal for $x$ and $z$ are less sensitive in the relevant \SIrange{100}{800}{\kHz} range.

\subsection{Imaging Sequence}
\label{sec:fields:Seq}

The MPI brain scanner superimposes two fast drive fields in the $x$- and $z$-direction and a dynamic selection field. The selection field is shifted along the $y$-axis, as explained in previous sections. Using and combining these constituents, it is possible to perform various imaging sequences. First of all, there are three different 2D sequences possible that drive the \ac{FFP} either along a 2D Lissajous trajectory with a frequency ratio of $\frac{75}{76}$ ($xz$-plane) or along Cartesian trajectories ($xy$- and $yz$-plane) with flexible densities. For an overview of common MPI sampling trajectories, we refer the reader to Knopp et al~\cite{knopp_trajectory_2008}. In the case of the $xz$-sequence, both \ac{DF}s are activated and a static \ac{FFP} field is generated at an eligible $y$-position, using the \ac{SFG}. For the other two cases, only one \ac{DF} channel is activated and the \ac{FFP} is swept dynamically along the $y$-direction. The repetition time of the $xz$-sequence is fixed and given by $T_\text{cycle}^{xz} = \SI{2.918}{\ms}$ and for the other two sequences, $T_\text{cycle}^{xy} = T_\text{cycle}^{yz} = \SI{0.25}{\s}$ if we set the $y$-sequence to \SI{4}{\Hz} (the default value for the scanner). We note, that the first generation of the scanner published in Graeser et al.~\cite{graeser_human-sized_2019} was 2D only and limited to the $xy$-sequence with a frame rate of \SI{2}{\Hz}.

In addition to the 2D sequences, it is also possible to apply fully 3D sequences. This is done by using both \ac{DF} channels and simultaneously generating a slowly varying selection field with an \ac{FFP} moving along the $y$-axis~\cite{kaethner_axially_2015}. The resulting sampling trajectory is a 2D Lissajous trajectory within the $xz$-plane that is slowly swept back and forth along the $y$-direction as visualized in \autoref{fig:Overview}. In this way, the volume is sampled plane by plane and the repetition time is derived from the $y$-sequence, resulting in $T_\text{cycle}^{xyz} = \SI{0.25}{\s}$ for a \SI{4}{\Hz} sequence.
The waveform of the currents applied to the selection field is chosen to be triangular, such that the sequence is always periodic, at constant velocity, and without sudden discontinuities, in order to avoid strong mechanical forces in the copper shield of the \ac{DFG} due to eddy currents. Furthermore, a triangular sweep minimizes the likelihood of \ac{PNS} and \ac{SAR} by the selection field. The triangular waveform provides a constant slew rate and results in an almost linear \ac{FFP} movement as the iron cores are not yet saturated. 
The chosen sequence type depends on the spatial dimensions and the requirements on temporal resolution. Within this work, we only show experiments that were performed using the 3D imaging sequence and therefore focus on this particular sequence in the following. The nominal \ac{FOV} captured by the 3D sequence assuming ideal magnetic fields is (\qtyproduct{84x85x67}{\mm}). The density of the FFP sampling trajectory within the $xz$-plane is inhomogeneous due to the sinusoidal excitation \cite{knopp_trajectory_2008} and the largest gap can be derived from the Lissajous node points \cite{kaethner_non-equispaced_2016}. For the applied Lissajous sequence, the resulting distance is \SI{1.819}{\mm}. The distance between slices in $y$-direction is equidistant with \SI{1.982}{\mm}. Both values are well below the expected spatial resolution between \SI{5}{\mm} and \SI{20}{\mm}, which ensures that the resolution is not limited by the sampling scheme.

\subsection{Operational Control}
\label{sec:field:OCU}
The task of the operation control is to implement the chosen imaging sequence by coordinating all scanner components used during a measurement. Next to the synchronous signal handling, an imaging sequence also requires several asynchronous tasks, such as enabling the various amplifiers, moving a calibration robot, updating the \ac{RPs} waveforms based on different imaging sequences or informing the \ac{SU} of an upcoming signal generation. These steps are handled by the open-source framework~\cite{hackelberg_mpimeasurementsjl_2023}, which can implement a variety of different measurement scenarios, including system matrix calibrations and magnetic field measurements~\cite{boberg_unique_2023}.

Depending on the type of measurement and sequence, the components and steps required may vary, but in general each measurement can be divided into three phases. During the setup phase, the amplifiers are disabled, the \ac{RPs} are neither transmitting nor receiving signals, and the resonance tuning heating is enabled. In this phase various components are prepared for the next imaging sequence, e.g. the calibration robot is moved or the \ac{RPs} waveforms are updated.
The next phase is the control phase. This is where the \ac{DF} amplifiers are enabled and the \ac{DF} waveform is sent down the transmit chain. During these transmissions, the feedback signal is analyzed at the frequencies of the excitation channels and the transmitted amplitude and phase are adjusted to the desired values from the selected sequence. When the deviation between the selected and observed \ac{DF} waveform is sufficiently small, the control phase ends. In measurement scenarios with frequent measurements, such as a system matrix calibration, the feedback signal from the previous measurement can be used instead to adjust the amplitude and phase.
The third phase is the active imaging phase. Here the resonance tuning heating is switched off to prevent possible distortions, both the \ac{SF} and \ac{DF} waveforms are transmitted, and the signal is received via the \ac{RPs} to be stored for further processing.

\section{Data Processing}
\label{sec:data}
\subsection{Image Reconstruction} 
\label{sec:data:recon}

The measured voltage signal of the $j$-th time frame $u_j(t,l)$ depends on the time $t\in\IR_+$ and the receive channel $l \in \set{1,\dots,L}$ where $L$ is the number of receive channels, i.e. $L=2$ in our case. The signal is sampled at time points $t_i = (i-1) / f_\text{sample}$, $i=1,\dots, K$ where $f_\text{sample} = \frac{125}{64}\,\text{MHz}$ and $K=484500$ for our 3D 4\,Hz sequence.
Prior to reconstruction, we apply the standard signal processing \cite{knopp_magnetic_2017}, which involves Fourier transformation, a frequency selection based on a bandstop filter in the receive chain filter (see above for details), as well as a filtering based on the SNR using different thresholds ranging from $3$ to $100$, resulting in $M$ frequency components.
For better readability of the following paragraphs, we express the resulting discrete signal as the vector $\mathbf{u}\in \IC^{M}$, omitting the frame index $j$.

The goal of image reconstruction is to recover an image $\mathbf{c} \in \IR_+^{N}$ that is discretized on a (3D) grid with $N$ voxels. The relationship between $\mathbf{u}$ and $\mathbf{c}$ is given by the linear system of equations $\mathbf{S}\mathbf{c} = \mathbf{u}$, where $\mathbf{S} \in \IC^{M\times N}$ is the system matrix that encodes the physical process from the particle magnetization progression to the system dependent measurement signal at the end of the receive chain.
Since the linear system is ill-conditioned and the measured data is disturbed by noise, we consider a regularized least-squares approach
\begin{equation}\label{eq:OptimizationApproach}
    \underset{\mathbf c\in\IR^N_+}{\text{argmin}}~\norm{\mathbf S \mathbf c - \mathbf u}_2^2~+~\lambda_{L^2} \norm{\mathbf c}^2_2~+~\lambda_{L^1} \norm{\mathbf c}_1 \,,
\end{equation}
where the first term ensures data consistency, the second term penalizes large solutions and prevents that the particle concentration vector $\mathbf{c}$ is fitted to the noise in the measurement $\mathbf{u}$, and the last term allows to penalize non-sparse solutions, which helps in reducing noise.
The optimization problem \eqref{eq:OptimizationApproach} is solved using the iterative Kaczmarz approach \cite{knopp_weighted_2010} using $L^2$-regularization as well as $L^1$-regularization~\cite{lieb_waveletbased_2021,dittmer_deep_2020}. The reconstruction has four parameters in total: $\lambda_{L^2}$, $\lambda_{L^1}$, the SNR threshold, and the number of iterations that are chosen based on visual inspection and experience.

The system matrix $\mathbf{S}$ encodes the applied imaging sequence and can be interpreted in two different ways. It can be interpreted either as a multi-patch dataset, where the entire 3D sequence is split into the individual 2D subsequences, or as a single-patch dataset. A patch refers to a subvolume that is moved by the dynamic selection field and consists of a single full \ac{DF} cycle.
The multi-patch approach is common when the applied selection field changes only in a step-wise fashion but it can also be applied when the selection field changes only slowly compared to the rapid \ac{DF} induced movement. This multi-patch reconstruction approach was used in the first generation of our perfusion imager \cite{graeser_human-sized_2019} and studied in more detail by Szwargulski et al~\cite{szwargulski_moving_2018}.
Alternatively, one can also interpret the entire imaging sequence as a single-patch dataset, which was considered by Werner et al.~\cite{werner_first_2017} for Cartesian 2D trajectories. The multi-patch approach has the potential advantage that it may allow to exploit shift-invariant sub-blocks within the system matrix, which can accelerate operations involving the system matrix $\pmb{S}$ within image reconstruction \cite{szwargulski_efficient_2019}. 
On the other hand, the single-patch approach can take field imperfections better into account and can also better prevent spectral leakage, which can be induced by non-periodic external signal contributions. Effectively, the key difference between both approaches is that the Fourier transform is applied to smaller signal snippets in the multi-patch case while the single-patch case applies the Fourier transform to the entire time signal.
Since the single-patch reconstruction is fast enough for our purposes, we use this approach in all experiments shown in this paper. For the system matrix analysis performed in \autoref{sec:res:SManalysis} we, however, consider both the multi-patch and the single-patch data interpretation since this gives much deeper insight into the system matrix structure.

\subsection{Perfusion Image Calculation}
\label{sec:data:perfcalc}

To evaluate perfusion experiments in \autoref{sec:res:perfExp}, the reconstructed data is processed to obtain the \acf{TTP}, \acf{MTT}, \acf{rCBF} and \acf{rCBV}. Post-processing includes 4 consecutive steps: (i) time framing, (ii) filtering and offset correction, (iii) threshold masking, (iv) parameter map calculation. After explaining these post-processing steps, we give the implemented definitions of the mentioned perfusion parameters. The definitions and the post-processing script that handles the data are based on the work by Mohn et al\cite{mohn_saline_2023}, which provides more details.

Step (i) selects the relevant reconstructed data by taking the time frames that include the entire first passage of the administered bolus. It starts with the injection ($5$\,s before bolus appearance), includes the administered bolus and stops after the passing levels to zero again. This step ensures that only relevant data is processed later, as shown in the left-most column of \autoref{fig:dynamic}. 
In step (ii), an appropriate filter type smooths the data for a more accurate peak detection and noise suppression. We avoided rectangular windows to reject ringing artifacts and selected a low-pass Hann-filter with a window-size of $10$ samples. The Hann-filter is applied voxel-wise on the Fourier transformed temporal data, which also shifts the concentration offset to zero by excluding the DC component. 
Afterwards in step (iii), a threshold mask reduces image noise by excluding any voxels with an intensity lower than $10$\,\% of the maximum value. This increases the readability of perfusion maps, by excluding irrelevant regions, e.g. outside (phantom) vessels.
A last post-processing step (iv), calculates the mentioned perfusion parameters based on the following definitions, which are also sketched in the top row of \autoref{fig:dynamic}. 
The reconstructed time-dependent solution of the entire volume in the \ac{FOV} is $\signatur{\tilde{\mathbf{c}}}{[0,T]}{\IR^N_+}$ with $N=n_x \times n_y \times n_z$ voxels. $\tilde{\pmb{c}}(t)$ describes the reconstructed volume of all measured time frames, that record the entire bolus administration. 

\paragraph{TTP:} We define the \acf{TTP} as the time elapsed between a chosen reference point (the bolus injection $t_0$) and the measured signal maximum of the bolus passing. The \textbf{TTP} $\in \IR^N$ is calculated element-wise for all voxel $n \in \{1,\dots,N\}$ via $\textup{TTP}_n = \textup{arg\,max}_t (\tilde{c}_n (t))$, where $\tilde{c}_{n}(t)$ is the concentration over time of the $n$-th voxel~\cite{fieselmann_deconvolution-based_2011}. 

\paragraph{MTT:} With \acf{MTT}, we refer to the measure of the average time interval that a particle or blood cell spends inside an organ or vessel and it strongly correlates with the \ac{FWHM} of the bolus concentration on passing (for low tissue perfusion)~\cite{kealey_user-defined_2004,ostergaard_principles_2005}. The time interval of the \ac{FWHM} was therefore selected as the \ac{MTT} $\in \IR^N$ in this work.

\paragraph{rCBF:} The \acf{rCBF} equals the highest positive gradient of the concentration over time $\tilde{c}_{n}(t)$, as in $\textup{rCBF}_n = \textup{arg\,max}_t (\frac{\textup{d}}{\textup{d}t} \tilde{c}_n(t))$, which is evaluated element-wise for all voxels $n$ to obtain \textbf{rCBF} $\in \IR^N$.

\paragraph{rCBV:} We derive the \acf{rCBV} data \textbf{rCBV} $\in \IR^N$  from an element-wise evaluation of the integral (\ac{AUC}) of the concentration $\tilde{c}_{n}(t)$ in the $n$-th voxel, over the time interval of the passing bolus ($\tilde{c}_{n}(t) > 0$) as in $\textup{rCBV}_n = \int \tilde{c}_n(t)\, \textup{d}t$. \\

\noindent Blood flow and volume are both calculated in a relative manner, due to a missing correct arterial input function (e.g. feeding phantom hose), which poses as a reference by providing the undisturbed flow (without perfusion).

\section{Experiments and Results}
\label{sec:results}

To evaluate the performance of the developed MPI brain scanner and its characteristics, several experiments were conducted. 
We start with a low-level evaluation of the scanner and first study the generated excitation and selection field by using appropriate field measurements. In the next step, we perform an analysis of the acquired system matrix $\pmb{S}$, which allows to derive and predict the imaging performance, independent of specific phantoms. Then, the imaging performance is analyzed at the phantom level using simple sensitivity and resolution phantoms and later also using an application-relevant dynamic perfusion phantom. Finally, the suitability of the scanner for multi-contrast imaging is demonstrated.

\subsection{Field Analysis}
\label{sec:res:fields}
\begin{figure}[!ht]
    \centering
    \includegraphics[width=\textwidth]{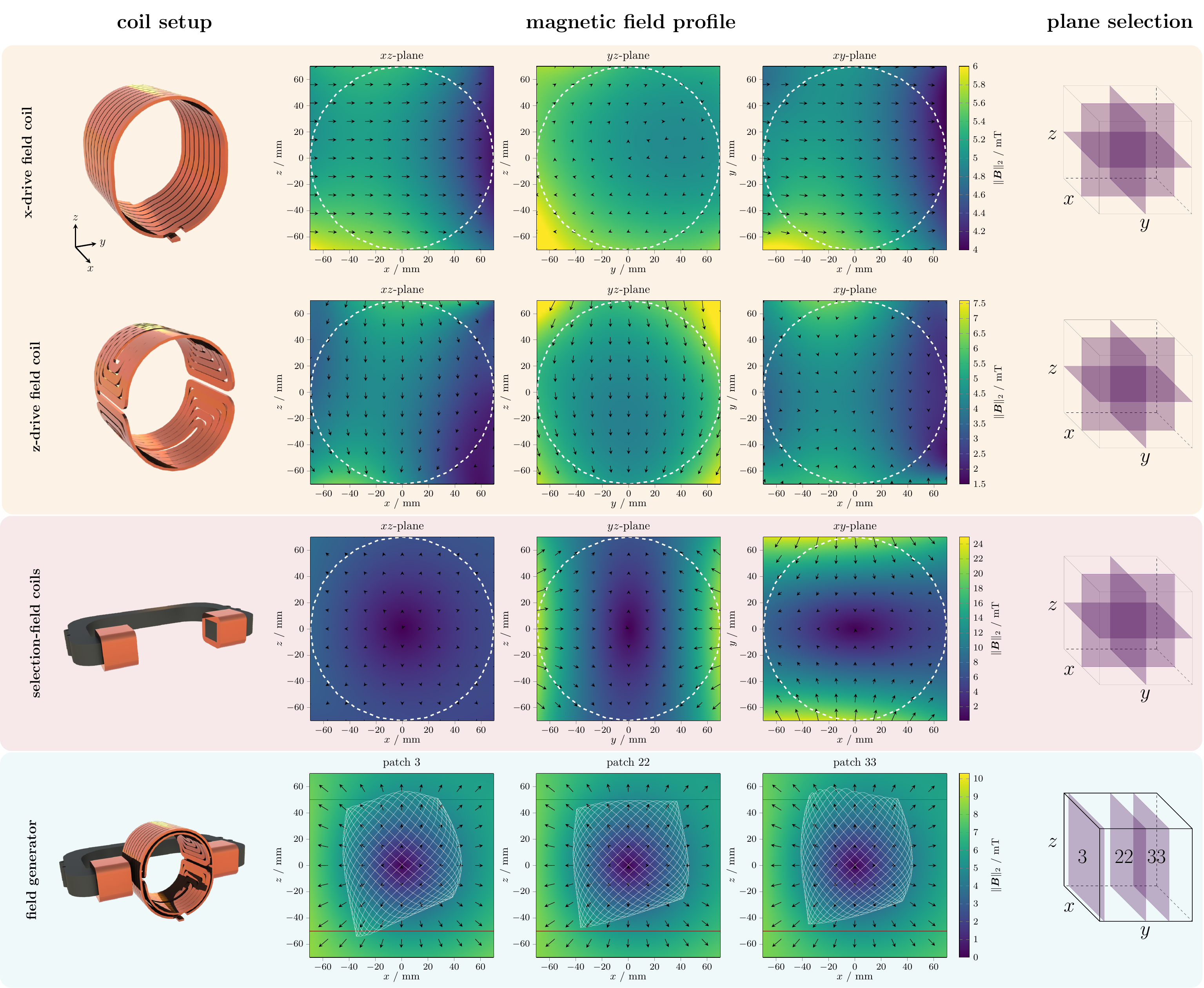}
    \caption{Field analysis of the magnetic field generating coils. On the left, the field generating coil setups are shown with the measured corresponding magnetic field profiles in the middle and an illustration of the respective planes on the right. The first two rows show the drive field of $x$ and $z$. The third row shows the selection field. All fields were measured on a sphere utilizing a \num{3}-axis field sensor and within the actual setup. For better visualization the copper shielding between selection-field coils and \ac{DFCs} is not shown.    
    In the bottom row, the \ac{DF} trajectory of the patches \numlist{3;22;33} is shown in the $xz$-plane, while the background features the corresponding selection field shifted in the $y$-direction. For visualization, the frequency ratio is adapted to $\frac{16}{15}$ and the system matrix \ac{FOV} is outlined in red.}
    \label{fig:fieldProfile}
\end{figure}

The image quality of an MPI scanner is closely linked to the \ac{FFP} trajectory, which in turn depends on the homogeneity of the drive and selection field. In the presented MPI scanner, the DFG is located in close proximity to the copper shielding of the selection-field generator (see \autoref{fig:Overview}). Hence, the drive field generates eddy currents inside the copper shielding, which in turn influences the field profile of the \ac{DFCs}. For the selection-field coils, the cross section area is relatively small compared to the distance between the coils, which leads to field inhomogeneities. 
To obtain the actual imaging trajectory, the field profiles of the drive and selection fields can be represented as a spherical harmonic expansion by measuring a few points on a sphere rather than using a Cartesian grid interpolation, as reported by Boberg et al~\cite{boberg_unique_2023}. A 1D transmit sequence at a defined \ac{DF} amplitude of \SI{5}{\milli\tesla} and \SI{4}{\milli\tesla} was used to measure the $x$- and $z$-drive field. The dynamic magnetic fields were determined using a calibration robot with a mounted custom 3D coil sensor, with \num{86} measurement positions of a spherical \num{12}-design~\cite{thieben_efficient_2022,hardin_library_nodate}. 
Due to the $x$-receive coil turns at the front of the \ac{DFG}, the \ac{FOV} center and thus the sphere center are shifted by \SI{23}{mm} in $x$-direction from the geometric \ac{DFG} center. In addition, these measurements were used to determine the fundamental \ac{THD} of the drive fields~\cite{shmilovitz_definition_2005}. The amplified and filtered \ac{DF} signal induced into the coil sensor exhibits a \ac{THD} of  \SI{0.0669}{\percent} and \SI{0.127}{\percent} for the $x$- and $z$-channel, respectively.
Regarding the selection field, the field of each coil was examined individually. Due to soft-iron induced saturation behavior, a list of \SIlist{2;4;6;8;10;12;14}{A} DC currents was set to measure the magnetic field at \num{36} positions of a spherical \num{8}-design~\cite{hardin_library_nodate} using a three-axis high-sensitivity Hall-effect sensor with a $3$-channel gaussmeter (Model 460, Lake Shore Cryotronics, Inc., USA). This measurement allowed the adjustment of the relationship between current, gradient strength, and \ac{FFP} position~\cite{boberg_unique_2023}.
In \autoref{fig:fieldProfile} the field profiles for the $x$- and $z$-drive fields are shown in the top rows. Field inhomogeneities become stronger towards the edges of the \ac{FOV}.

The selection field for identical \SI{10}{A} currents in both coils is shown in the third row of \autoref{fig:fieldProfile}. The largest gradient is observed in the $y$-direction, reaching \SI{0.24}{\Tpm}. In comparison, the gradients in the $x$- and $z$-direction are half as strong, measuring \SI{-0.12}{\Tpm}. 
In the bottom row of \autoref{fig:fieldProfile}, the 2D Lissajous trajectory is shown in the $xz$-plane for defined selection-field offsets of patch \numlist{3;22;33}. The trajectory is sampled by marking the FFP after superposition of all three magnetic fields for certain time points. For visualization the density of the trajectory is adapted by changing the frequency ratio to $\frac{16}{15}$.  In the background, the selection field of each patch is shown in the $xz$-plane. 
From the measurements, the calculated 3D \ac{DF} FOV spans a volume of \qtyproduct{83x80x73}{\milli\meter}.

\subsection{System Matrix Analysis}
\label{sec:res:SManalysis}

In the next step, the imaging performance of the scanner is analyzed by studying a measured system matrix, which will later also be used for image reconstruction. The system matrix was acquired  using a robot-based approach with a cubic \SI{200}{\micro L} $\delta$-sample filled with \textit{Resotran}\,(b.e.imaging GmbH, Germany) in a concentration of \SI{8.5}{\mg\of{Fe}\per\ml} (\SI{152}{\mol\per\L}) to prevent magnetic dipole-dipole interactions~\cite{lowa_concentration_2016}. The sample was shifted to \numproduct{15 x 15 x 11} positions covering a volume of \qtyproduct{140 x 110 x 100}{\milli\meter} and at each position one full 3D sequence was applied. For later background subtraction and SNR analysis, \num{12} empty measurements after each $xy$-plane were performed by moving the sample with the calibration robot outside the scanner bore. During the acquisition, the \ac{DF} feedback was tracked and the observed amplitudes and phases showed a standard deviation of below \SI{0.4}{\percent} over the 2491 measurements.  All signals are considered in frequency space, which is common in MPI since it allows for direct filtering of interfering signals like the signal induced by the drive field.

\begin{figure}[!ht]
    \centering
    \includegraphics[width=1.0\textwidth]{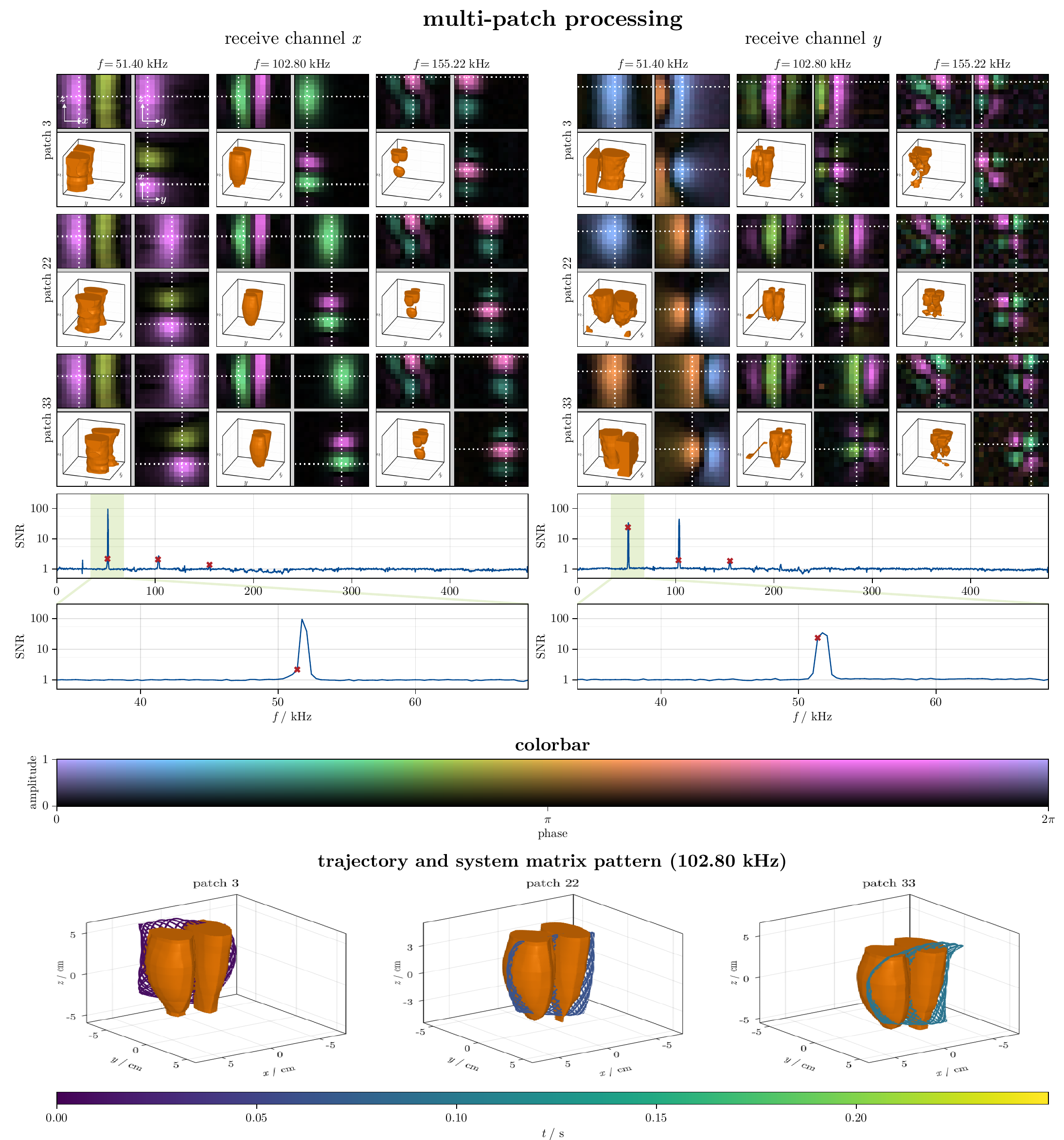}
    \caption{Measured system matrix considering a multi-patch processing. The upper part of the figure shows selected frequency components for the $x$-(left) and the $y$-(right) receive channel. Each 3D frequency component is visualized by plotting three orthogonal slices and an iso-surface rendering of the magnitude. The slice position is indicated as a white dotted line. The frequency component is shown for three different patch positions ($p\in \{3,22,33\}$). The middle part of the figure shows the SNR of the system matrix rows as a function of frequency. The SNR is visualized in a hierarchical fashion using two different nested frequency ranges, which are indicated by light green boxes. The lower part of the figure shows, for one selected frequency component, the 3D iso-surface rendering in combination with the actual FFP trajectory for the considered patch. Here, the color encodes the time within the full imaging sequence.}
    \label{fig:SMMultiPatch}
\end{figure}

The measured system matrix can be interpreted in two different ways. First, since the FFP movement induced by the selection-field generator is very slow, the data can be interpreted as a multi-patch dataset, where the 2D Lissajous trajectory ($xz$-plane) is slowly shifted to $M_y =85$ positions along the $y$-axis. One can thus expect that the system-matrix patterns are just shifted in space, which was shown for idealized magnetic fields by Szwargulski et al.~\cite{szwargulski_efficient_2019} and Boberg et al~\cite{boberg_generalized_2020}. The measured system matrix considering this multi-patch processing is illustrated in \autoref{fig:SMMultiPatch}. Shown are frequency patterns -- i.e. system matrix rows reshaped on the 3D grid -- at frequencies $f_k^\text{MP} = k \Delta_{f}^\text{MP}$ where $\Delta_f^\text{MP} = \frac{1}{T_\text{cycle}^{\text{MP}}} \approx \SI{342.654}{Hz}$ is the frequency spacing derived from the 2D Lissajous sequence length $T_\text{cycle}^{\text{MP}} \approx$ \SI{2.918}{ms} and $k\in\IN$ is the frequency index. The index $k$ can be expressed using the mixing factors $m_x$ and $m_z$ as $k_{m_x,m_z} = m_x M_x + m_z M_z$ where $M_x = f_x^\text{DF} T_\text{cycle}^{\text{MP}} = 76 $ and $M_z = f_z^\text{DF} T_\text{cycle}^{\text{MP}} = 75$, see Rahmer et al\cite{rahmer_analysis_2012}. The figure shows on the left selected frequency patterns ($(f,m_x,m_z) \in \{ (\SI{51.4}{kHz}, 2, 0), (\SI{102.8}{kHz}, 4, 0), (\SI{155.22}{kHz}, 3, 3)\}$ of the $x$-receive chain and on the right of the $y$-receive chain. Each pattern is visualized using an iso-surface rendering (lower left) and three orthogonal slices (upper left: $xz$, upper right: $yz$, lower right: $xy$) where the dotted line indicates the slice position. The complex-valued colormap encodes the amplitude in the saturation and the phase in the color \cite{albers_modeling_2022}. In the middle part of the figure, the mean SNR over all patches is shown as a function of frequency. To illustrate both, the global and the local SNR progression is plotted for different frequency ranges. The lower part of the figure shows a 3D iso-surface rendering of frequency component \SI{102.80}{\kHz} combined with the actual FFP trajectory derived from the measured fields for the three considered patches.
The measured system matrix shows the expected wave-like structure in which the number of extrema depends on the mixing factors\cite{szwargulski_influence_2017}. For $m_x=3,m_y=3$ one can see oscillating patterns in $x$- and $z$-direction, which shows that the sequence spatially encodes both directions. When considering the patch movement, it is clearly visible that the frequency patterns are shifted when the FFP sweeps slowly in $y$-direction. However, slight distortions of the patterns are observable, which are caused by field imperfections and violate true shift-invariance of the system. 

\begin{figure}[!ht]
    \centering
    \includegraphics[width=0.96\textwidth]{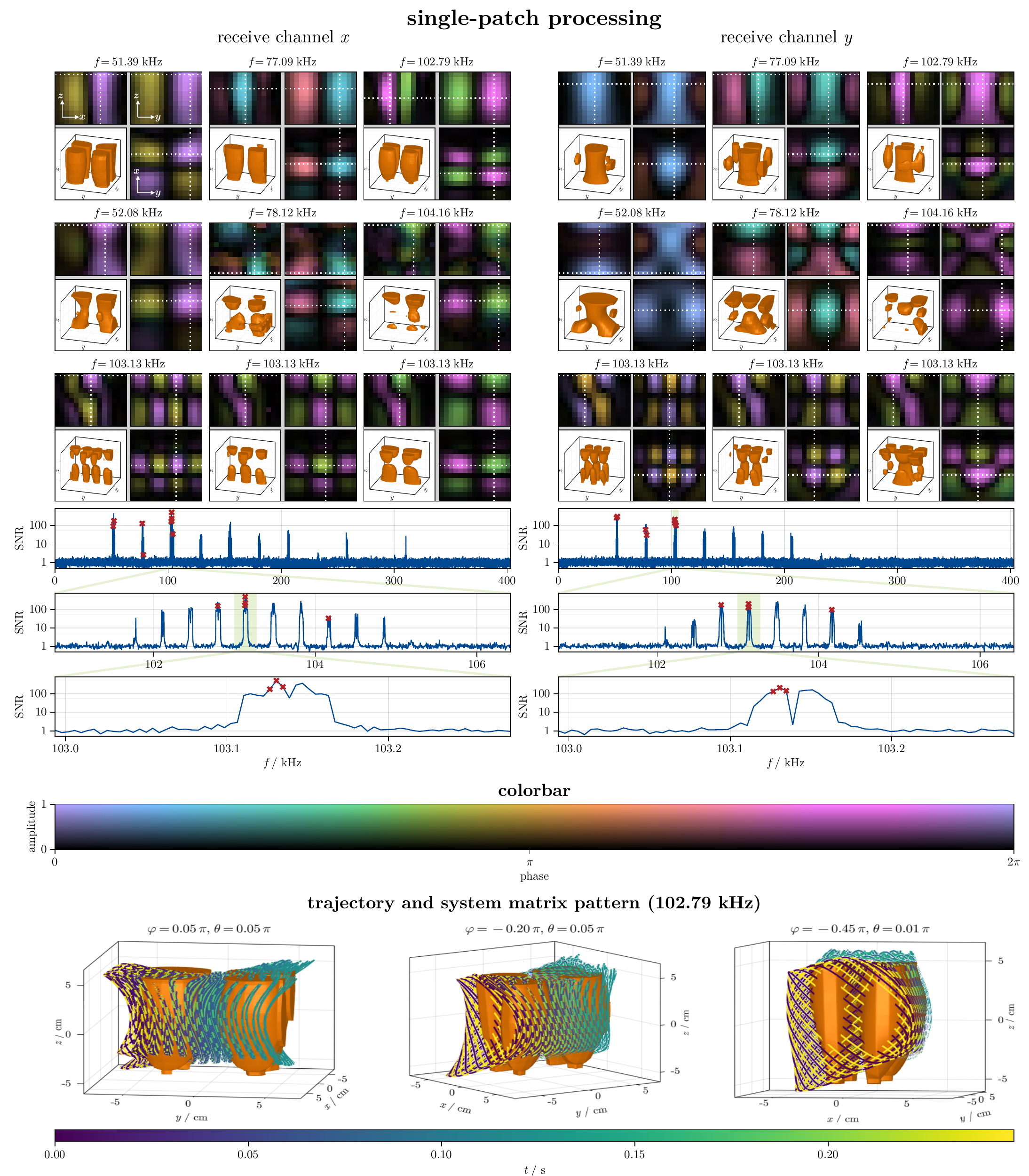}
    \caption{Measured system matrix considering a single-patch processing. The upper part of the figure shows selected frequency components for the $x$-(left) and the $y$-(right) receive channel. Each 3D frequency component is visualized by plotting three orthogonal slices and an iso-surface rendering of the magnitude. The slice position is indicated as a white dotted line. In the center, the SNR of the system matrix rows are shown as a function of frequency. The SNR is visualized in a hierarchical fashion using three different nested frequency ranges, which are indicated by a light green box. The lower part of the figure shows for one selected frequency component the 3D iso-surface rendering in combination with the actual FFP trajectory viewed from three different angles. Here, the color encodes the time within the imaging sequence.}
    \label{fig:SMSinglePatch}
\end{figure}

Next, we consider the system matrix not as a multi-patch dataset but as a single-patch dataset. This is possible because there are no temporal gaps in the sequence and thus the sequence can be considered to be 3D with fast FFP movement in the $xz$-plane and slow FFP movement in the $y$-direction. In turn, the sequence time increases to $T_\text{cycle}^\text{SP} = M_y T_\text{cycle}^\text{MP} \approx \SI{248.06}{\ms}$ and the frequency spacing in turn decreases to \SI{4.031}{\Hz}. This means that the single-patch spectrum contains much more frequencies (factor $M_y$) but that the patch-encoding dimension is lost. This is illustrated in \autoref{fig:SMSinglePatch}, which now shows in the SNR plot a signal with two levels of sub-bands. The SNR is slightly higher compared to the multi-patch case since an implicit averaging takes place when applying the Fourier transform to the longer time interval. In the single-patch case, the signal occurs at frequencies $f_k^\text{SP} = k \Delta_{f}^\text{SP}$ and the frequency index $k$ can now be expressed as $k = m_y + M_y (m_x M_x + m_z M_z)$ where $m_y$ is now a new mixing factor that encodes the finest level of frequency shifts. The upper part of \autoref{fig:SMSinglePatch} shows selected frequency components for various mixing factors sampling different sub-bands of the frequency space. Again, the expected wave-like patterns are visible, but this time the patterns do not only surround the trajectory plane, but cover the entire FOV and also show oscillating structures in the $y$-direction. The lower part of the figure shows frequency component \SI{102.79}{\kHz} of the $x$-receive chain in combination with the full 3D FFP trajectory from three different angles. One can again see that both the system matrix pattern and the trajectory are rotated within the $xz$-plane due to field imperfections.

\begin{figure}[!ht]
    \centering
    \includegraphics[width=1.0\textwidth]{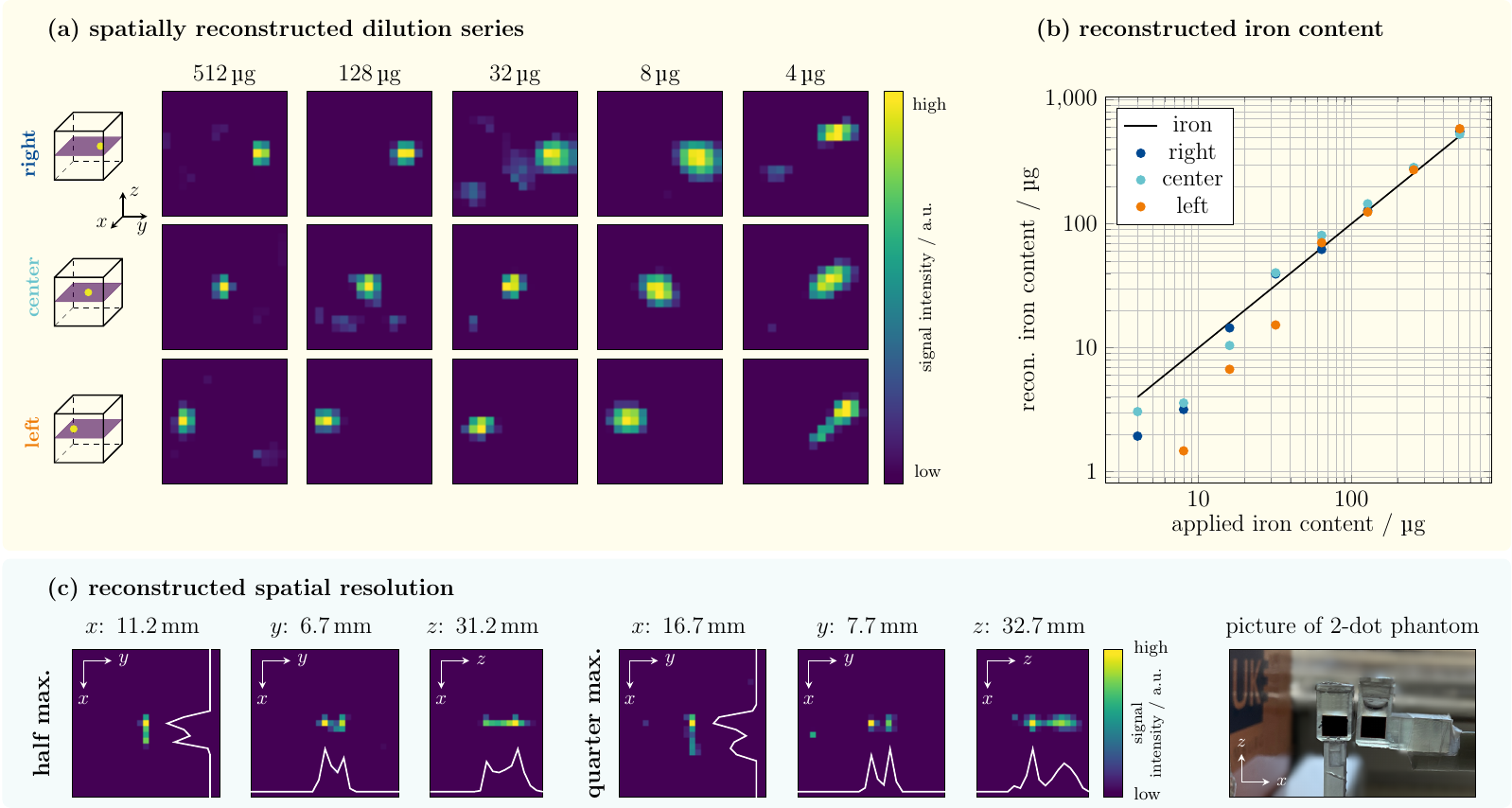}
    \caption{Sensitivity study and spatial resolution experiments. Reconstruction results for the sensitivity of descending iron contents inside a \SI{200}{\micro\litre}  Eppendorf tube with \SI{50}{\micro\litre} volume of tracer are shown in the central $xy$-plane for multiple positions in (a). For each iron content and each position, the quantitative reconstructed iron content is mapped to the applied iron content in (b). For the spatial resolution experiment using \textit{Resotran} in (c), two  $\delta$-samples (\SI{200}{\micro\litre} each, \SI{8.5}{\mg\of{Fe}\per\ml} (\SI{152}{\mol\per\L}) were used. An edge-to-edge distance was achieved by placing one sample in the center of the \ac{FOV} while the other was moved by the calibration robot to defined positions with increasing distance (\SI{0.5}{mm} steps).
    Images were recorded using the 3D sequence with dedicated $x$- and $y$-receive channels and reconstructions are based on a single frame (one average) with an interpolated grid \numproduct{25x25x19}. The spatial resolution was considered successful, when the profile line in the middle between the two samples dropped below the half maximum (white line). Results for a quarter maximum criterion are shown as well.}
    \label{fig:sensitivity_spatial}
\end{figure}

\subsection{Sensitivity and Spatial Resolution Experiments}
\label{sec:res:sensiresol}

To determine the sensitivity of the MPI scanner, we implemented the protocol using three different spatial positions developed in Graeser et al~\cite{graeser_towards_2017}. First, a dilution series of the tracer \textit{Resotran} was prepared with \SI{50}{\micro\litre} samples inside \SI{200}{\micro\litre}  Eppendorf tubes and ascending iron mass between \SI{4}{\micro\g} and \SI{512}{\micro\g}. Eight samples were measured at three spatial locations each, positioned along the $y$-axis. This facilitates a distinction between the sample signal and reconstruction artifacts. The reconstruction results are shown in \autoref{fig:sensitivity_spatial}\,(a) for several iron masses. The spatial position can be resolved down to \SI{8}{\micro\g} iron. At \SI{4}{\micro\g} iron, a blurred dot can be seen, but with a large positioning error, which is consequently a failed reconstruction.
For quantitative analysis, a post-processing step sums the reconstructed particle concentration within the mask around the sample position and multiplies it by the iron mass of the system matrix $\delta$-sample. The results are shown in \autoref{fig:sensitivity_spatial}\,(b). At higher iron mass, the reconstructed iron content matches the applied iron content. For lower iron mass, below \SI{32}{\micro\g}, the reconstructed particle concentration becomes smaller than expected.

To assess the spatial resolution, a \SI{200}{\micro\litre} $\delta$-sample of \textit{Resotran} with \SI{8.5}{\mg\of{Fe}\per\ml} (\SI{152}{\mol\per\L}) iron was placed in the \ac{FOV}-center. A second, identical $\delta$-sample was mounted on a rod, which could be positioned by a calibration robot. The latter was moved directly next to the centered sample. After imaging using the 3D sequence, the edge-to-edge distance was increased by \SI{0.5}{mm} until reconstruction could discriminate the two samples. This procedure was performed for all three main axes. For better reconstruction results, the system matrix grid was interpolated to \numproduct{25x25x19}. The image signal was summed up over 3 voxel inside a mask around the direction of interest to generate profile lines. The reconstruction was defined to be resolved if the profile line dropped below half the maximum in the middle between the two samples.  
In \autoref{fig:sensitivity_spatial}\,(c), the resolved reconstructions are shown for a half and quarter maximum criterion for all the directions. Additionally the profile lines are shown in the reconstructed images.
With the half maximum definition, the best spatial resolution is found in the $y$-direction with \SI{6.7}{mm}, followed by \SI{11.2}{mm} in $x$- and \SI{31.2}{mm} in $z$-direction.

\subsection{Dynamic Perfusion Experiments}
\label{sec:res:perfExp}
\begin{figure}[!ht] 
    \centering \includegraphics[width=1.0\textwidth]{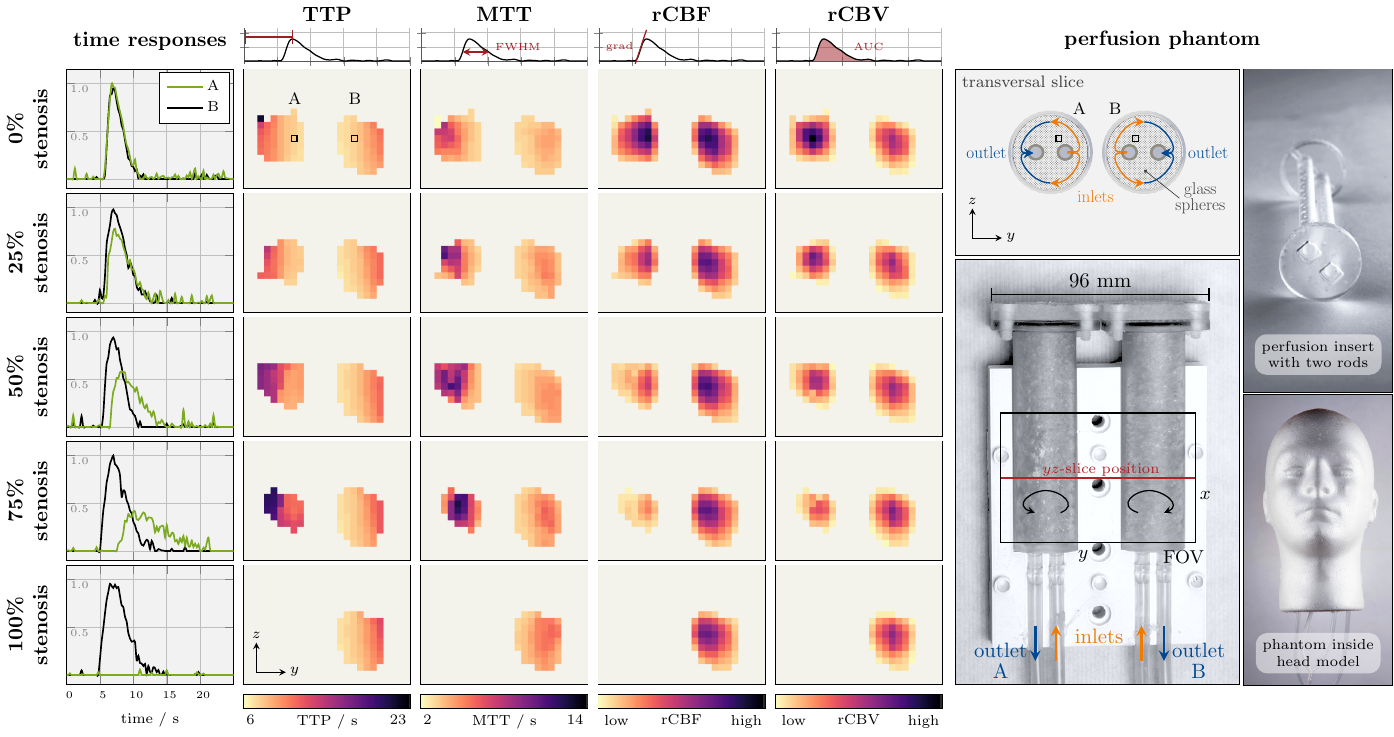}
    \caption{Perfusion phantom for fast dynamic imaging. The volumes were acquired with the 3D sequence. Two cylinders represent the left and right hemisphere of the brain and each \SI{50}{\milli\litre} cylinder is filled with \SI{1}{mm} diameter glass spheres. A perfusion insert with two perforated rods sits inside the cylinder, one connected to the feed (inlet, orange) and the other as a drain (outlet, blue). They contain several holes on opposite facing sides, to mimic tissue perfusion, and are shown in the pictures on the right. The expected flow is visualized for the transversal $yz$-slice, that is depicted throughout the reconstructed perfusion images. Different levels of stenosis in 25\,\% steps were simulated by using two independent peristaltic pumps, connected to one output each. The suction was regulated to match a flow rate of \SI{100}{\ml\per\minute} for the healthy brain half (B, on the right in each case). For the highlighted voxels A and B, the time responses of the normalized concentration are shown. 
    The bolus contained \SI{100}{\micro\litre} of pure \textit{Resotran} (\SI{28}{\mg\of{Fe}\per\ml}).
    From the data of the time responses, the \acf{TTP}, \acf{MTT}, \acf{rCBF}, and \acf{rCBV} perfusion maps were calculated. The \ac{rCBF} and \ac{rCBV} were normalized to the maximum value in the imaging volume. All time data were shifted to the arrival time of the bolus.}
    \label{fig:dynamic}
\end{figure}

For the analysis of volumetric imaging at high temporal resolution, we used a flow phantom filled with glass spheres and bolus injections of \textit{Resotran}. The phantom in this experiment is similar to the one developed by Graeser et al~\cite{graeser_human-sized_2019}. It consists of two tubes with a volume of \SI{50}{\ml} each, filled with \SI{1}{\mm} diameter glass spheres to simulate capillaries within the tissue. Two rods are placed inside the tubes with evenly distributed holes facing opposing directions, as shown in \autoref{fig:dynamic} on the top right. The tubes represent the two hemispheres of the brain and each tube is connected to a peristaltic pump, which uses suction to deliver an adjustable flow rate that is independent of the other hemisphere. Five experiments were conducted with different average flow rates: While the flow rate in the right hemispheres remained constant at \SI{100}{\ml\per\minute},  \SIlist{0;25;50;75;100}{\percent} of stenosis was obtained for the left hemisphere by reduced flow rates. 
Prior to experiments, the overall flow rates of outlet A and B were precisely matched by independent reference experiments to compensate for pressure differences.
Each experiment was measured over 150 frames (\SI{37.5}{\s}) and a \SI{100}{\uL} bolus of pure \textit{Resotran} (\SI{28}{\mg\of{Fe}\per\ml}) was administered.
For imaging, the same 3D sequence was used as for the sensitivity and resolution experiments, which is described in detail in \autoref{sec:fields:Seq}. During the perfusion measurements the observed \ac{DF} showed a standard deviation below \SI{0.3}{\percent} over the \num{150} consecutive frames without additional control steps. The data processing is divided in a reconstruction step, described in \autoref{sec:data:recon}, and a post-processing step that is build upon the reconstruction results and yields different perfusion parameters, as described in detail in \autoref{sec:data:perfcalc}. For the reconstruction, the system-matrix grid size was interpolated to $N = n_x \times n_y \times n_z = 25\times25\times18$ voxels. The reconstruction results revealed that the relevant data for the post-processing step ended after $t=25\,s$.

The results are illustrated in \autoref{fig:dynamic}. They indicate that fast dynamic imaging is feasible and different levels of stenosis can be detected and distinguished through calculated perfusion maps. Throughout the referenced figure, a transversal 2D slice is shown and its relative position is indicated in the picture of the phantom on the right side. This slice is exemplary for the entire 3D tomogram. The time response graphs in the left-most column show reconstructed data for voxel A and B, prior to Hann-filtering. Post-processing produced the 3D \ac{TTP}, \ac{MTT}, \ac{rCBF} and \ac{rCBV} maps and the definitions of each perfusion parameter is sketched above its column. In the case of a stenosis, the flow-suppression is visualized by a delayed signal peak in the \ac{TTP} map and an increasing transit time in the \ac{MTT} map (dark colors). Relative blood flow and volume decrease with the severity of the stenosis, which is revealed by lighter colors and a smaller area of outlet A. The decreasing area can be attributed to the threshold mask, which eliminates values below \SI{10}{\percent} of peak intensity. Slight differences of \ac{MTT} or \ac{rCBV} in the case of equal flow rate (\SI{0}{\percent} stenosis) are caused by variations of the phantoms, their filling and air cavities, which all influence the internal flow.
The entire set of 4 perfusion parameters reveal not only changes due to the stenosis, but also the increments of these changes become visible in each perfusion map.

\subsection{Multi-Contrast Experiments}
\label{sec:res:MC}
To demonstrate the ability of multi-contrast imaging within the MPI brain scanner, a simple 2-dot phantom and two different tracers were chosen. The $\delta$-samples of \textit{Resotran} and \textit{synomag} (micromod Partikeltechnologie, Germany) contained \SI{200}{\micro\litre} each with an iron concentration of \SI{8.5}{\mg\of{Fe}\per\ml} (\SI{152}{\mol\per\L}). For this proof-of-concept, two system matrices were recorded in the $xy$-plane on a \qtyproduct{15x15x1}{} grid with a system matrix FOV of \qtyproduct{140x110}{\mm} in $x$- and $y$-direction with \ac{DF} amplitudes of \SI{5}{\mT} in $x$ and \SI{4}{\mT} in $z$. The $\delta$-samples were used for the imaging experiments, mounted on a 3D printed platform and inserted by a calibration robot into the center of the scanner. The imaging sequence was the same as in all other experiments above. A single frame was recorded and reconstructed, using background subtraction. Reconstruction followed the same protocol as described in \autoref{sec:data:recon}, with the only difference that two measured system matrices are passed to the Kaczmarz-solver to separate the signal contributions of each tracer sample~\cite{rahmer_first_2015}.

\begin{figure}[!ht] 
    \centering \includegraphics[width=0.6\textwidth]{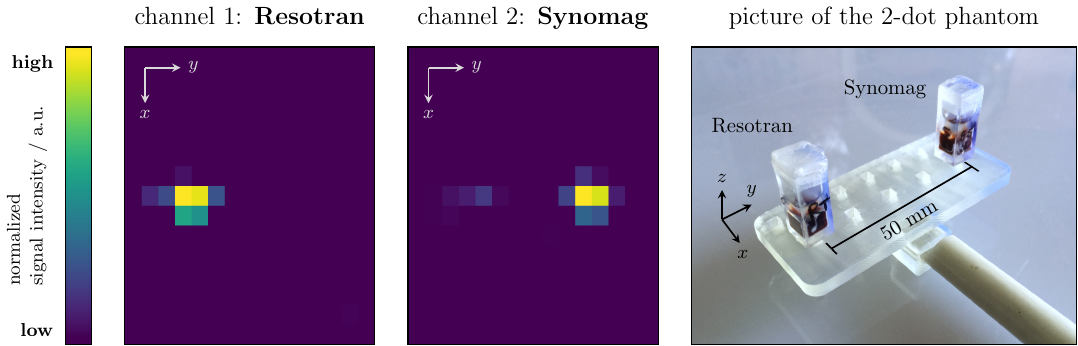}
    \caption{Multi-contrast experiment, using \textit{Resotran} and \textit{synomag}. The two  $\delta$-samples (\SI{200}{\micro\litre} each, \SI{8.5}{\mg\of{Fe}\per\ml}) were separated by a distance of \SI{50}{\mm}. The image was recorded using the 3D imaging sequence with $x$- and $y$-receive channels and the reconstruction is based on a single frame (one average). The FOV size is \SI{140}{\mm} in $x$-direction and \SI{110}{\mm} in $y$-direction. The signal intensity is normalized for each channel (system matrix) individually, however, the MPI tailored tracer \textit{synomag} has a stronger particle response. Overall, separation of the tracer samples was successful.}
    \label{fig:multicontrast}
\end{figure}

The resulting concentrations for both reconstruction channels are shown in \autoref{fig:multicontrast}, along with a picture of the phantom. 
Both $\delta$-samples are correctly depicted in their respective channel and spatially separated.
In the \textit{synomag} channel, a residual signal from the \textit{Resotran} sample can be observed. However, this channel leakage is small compared to the intrinsic signal of the \textit{synomag} sample.

\section{Discussion}
\label{sec:discussion}
In this study, we investigated a human-sized \ac{MPI} brain scanner designed to provide 3D imaging with high spatial resolution, \SI{4}{\Hz} temporal resolution, and with a focus on enabling the detection of neurovascular diseases.
We achieved a resolution of $12$\,mm in $x$-, $7$\,mm in $y$-, and $31$\,mm in $z$-direction using the clinically approved Ferucarbotran tracer \textit{Resotran}. Moreover, our scanner was able to detect an iron dose down to \SI{8}{\micro\gram}. Perfusion experiments were performed and images reveal that $5$ levels of stenosis (25\% increments) can be differentiated. Furthermore, we were able to discriminate \textit{Resotran} and \textit{synomag} within the same \ac{FOV} in multi-contrast images. 
Instrumentation safety was a major focus in the design and implementation of the transmit- and receive chains, as well as obtaining a 3D imaging sequence that samples a \SI{480}{\ml} volume. 
The scanner can potentially be used in an \ac{ICU} due to its manageable total power consumption of less than \SI{4}{\kW} and its robustness to electromagnetic interference in an unshielded environment.

\subsection{Imaging Capabilities}
The spatial resolution of an \ac{MPI} system is an elaborate interaction of gradient-field strength and excitation~\cite{rahmer_signal_2009}, imaging sequence~\cite{knopp_trajectory_2008}, magnetization response of particles~\cite{rahmer_signal_2009}, and the receive chain~\cite{szwargulski_influence_2017}. 
All measurements were obtained at an excitation amplitude of \SI{5}{\mT} in $x$- and \SI{4}{\mT} in $z$-direction, with a gradient strength of \SI{0.24}{\tesla\per\m} in $y$-direction (measured in the \ac{FOV} center).
Despite being orthogonal to both \ac{DF} directions, the best spatial resolution of $7$\,mm was found in the $y$-direction.
It benefits from a stronger gradient strength, the dedicated receive coil and the multi-patch excitation sequence.
The $x$-direction benefits from being aligned with one of the \ac{DF} directions and a dedicated receive coil, but we observe only about half the gradient strength. This culminates to a spatial resolution of $12$\,mm.
Due to the lack of a dedicated receive coil and the same gradient strength as in $x$, the spatial resolution in the $z$-direction is inferior with $33$\,mm, although it aligns with an \ac{DF} direction. 
In terms of sensitivity, compared to the previous version of the head scanner~\cite{graeser_human-sized_2019}, we reached a detection limit of \SI{8}{\micro\g\of{Fe}}, whereas the old limit was \SI{2}{\micro\g\of{Fe}}.
However, this does not imply that the redesign has a lower sensitivity by a factor of $4$, since a direct comparison does not prove to be feasible due to the differences in the experimental configuration and particle system. Reasons for an expected reduction of the sensitivity and spatial resolution are the topology of the imaging sequence that samples a 3D volume in the current version instead of a 2D slice. On the other hand, our new imaging sequence is only half as long and the signal intensity of the tracer \textit{Resotran} is lower than \textit{perimag} (micromod Partikeltechnologie, Germany), which implies that the actual sensitivity difference is lower.

The dynamic perfusion experiment takes full advantage of the temporal resolution of 4\,Hz and its 3D imaging capability.
Setting realistic flow-rates of \SI{100}{\ml\per\minute}, we were able to visualize a bolus passing through our perfusion phantom~\cite{chu_flow_2000}. 
In terms of temporal resolution, we calculated 4 different perfusion parameters in $5$ different levels of stenosis.
The spatial resolution of our scanner is sufficient to separate left and right brain hemisphere in all experimental settings.
However, the resolution required to image the cerebral vasculature (brain angiography) is not yet achieved. 
In a clinical scenario, the size of this bolus would need to be increased, due to the amount of blood delivered to the brain from an intravenous bolus (fraction of 15 to 20\,\%~\cite{xing_distribution_2017}). Assuming a fraction of 20\,\%, the amount of administered boli would be restricted to about 3 per prefilled vial (\SI{1.4}{\ml}), although limitations depend on body mass and metabolism~\cite{reimer_ferucarbotran_2003,southern_commentary_2018}.
A plausible long-term monitoring scenario would require a larger number of administrable boli, to facilitate interventions and evaluate treatment success afterwards. To increase the number of boli, we could reduce the bolus concentration, as the ability of perfusion parameter calculation is likely possible for lower signal strengths of the time response curves. A specifically tailored \ac{MPI} tracer, which could be clinically approved in the future, would increase the measurement signal leading to a lower iron dose per bolus. 
A complementary approach could make use of negative boli to increase the total number of administrable boli~\cite{mohn_saline_2023}.

Multi-contrast imaging was performed utilizing the clinically approved tracer \textit{Resotran} and the \ac{MPI} tailored tracer \textit{synomag}. For this proof-of-concept, we used \SI{200}{\micro\litre} samples at a concentration of \SI{8.5}{\mg\of{Fe}\per\ml} (\SI{152}{\mol\per\L}) and a single measurement without averaging. We observe minor channel leakage in the multi-contrast tomogram, which is typical for multi-contrast imaging~\cite{rahmer_first_2015}. Yet, localization and discrimination of both samples was successful and has the potential to provide additional information like temperature in hyperthermia~\cite{stehning_simultaneous_2016,chandrasekharan_using_2020}.

Common to the imaging sequence of the current and old scanner version, is the combination of a slow selection-field shift in $y$-direction with fast orthogonal \ac{DF} excitation. The old scanner used a 1D \ac{DF} in $x$-direction, which results in an imaging trajectory that samples a 2D \ac{FOV} in the $xy$-plane at \SI{2}{\Hz}. In contrast, the current design uses a Lissajous type \ac{DF} excitation in $xz$-plane, sampling a 3D \ac{FOV} (\SI{480}{\ml}; the volume of an adult human brain is about \SI{1200}{\ml}~\cite{parent_carpenters_1996}) at \SI{4}{\Hz}.
We note that both sampling trajectories are redundant by a factor of two since they contain a sweep along the positive and negative $y$-direction. This yields the potential to increase the frame rate by a factor of two when reconstructing the two halves independently, as has been proposed by Brenner at al~\cite{brenner_recovering_2023}.
A major improvement of the redesign is the extension of the \ac{FOV} to a 3D volume, sampled at twice the frame rate, which is a leap towards imaging the entire human brain.
Prototypes of multi-coil iron-core selection-field generators are in development, that further enlarge the \ac{FOV} and extend the space of feasible imaging sequences~\cite{foerger_flexible_2023,gdaniec_fast_2017}.

\subsection{Device Safety}

The presented scanner in this work uses the same selection-field yoke as the prior version by Graeser et al.~\cite{graeser_human-sized_2019}, however, all other major hardware components (the \ac{DFG}, \ac{ICN}, \ac{HCC}, transmit and receive chain) were redesigned for this scanner upgrade to achieve 3D imaging and focus on human safety for the first time. To this end, the limitation of high voltages in the vicinity of the imaging volume and the implementation of an independent \acl{SU} were key concerns. 
On a path towards human trials, safety concerns and regulations regarding the \ac{SAR}, \ac{PNS}, and conductors in patient proximity influence the scanner design, imaging sequence and limit maximum magnetic field strengths. 
The \ac{SAR} limitations in the head are given with \SI{3.2}{\W\per\kilo\gram} for the chosen \ac{DF} frequencies in the \si{\kHz} range~\cite{bohnert_effects_2010,chandrasekharan_using_2020}. However, for sinusoidal electric fields below \SI{100}{\kHz}, \ac{PNS} concerns prevail and are the limiting factor for alternating field strengths~\cite{dalziel_effect_1950,chatterjee_human_1986,reilly_maximum_1998}. 
Heeding \ac{PNS} limitations, the maximum \ac{DF} amplitude was set to \SI{5}{\milli\tesla}, which is realistic for human trials~\cite{saritas_magnetostimulation_2013,ozaslan_pns_2022}. 
Compared to \ac{MRI} gradients, the maximum slew rate of the dynamic selection field ($\approx$~\SI{24}{\tesla\per\meter\per\s}) is very low and below the risk threshold for \ac{PNS}~\cite{davids_peripheral_2023,zhang_peripheral_2003}. 
In \ac{MRI}, all ferromagnetic components must be excluded from entering the scanner, including pacemakers, mechanical ventilators and oxygen cylinders~\cite{porte_imaging_2009}. Potentially, this does not apply to our scanner, due the confinement of high magnetic fields within the head region only. Further investigations on specific device compatibility are required.

Electrical safety regarding shock and discharge, are mainly addressed by reducing the inductance of the \ac{DFG}. The necessary power to obtain the same field strength is thus provided by a high coil current, which creates a maximum voltage of \SI{535}{\V} (\SI{240}{\A} at \SI{5}{\mT}, \SI{14.4}{\micro\henry}) that is a reduction of a factor of about 4 compared to the previous version of the \ac{DFG}~\cite{graeser_human-sized_2019}. 
To achieve the capability to conduct around \SI{300}{\A} with Litz wire,
we used Rutherford wire parallelization in both \ac{DF} coils, which also minimizes the proximity effect. Moreover, the polyamid housing of the \ac{DFG} provides dielectric isolation and increases breakthrough voltage between \ac{DFG} components and the patient. 
Due to the \ac{ICN}, the entire \ac{HCC} obtains a floating potential, which implies that touching a single exposed point of the circuit is not a hazard, because residual leakage currents towards ground, e.g. caused by capacitive coupling, are below \SI{20}{\mA}~\cite{burgess_electrical_2019}. Thus, the insulation would have to fail at two separate points simultaneously, and both would have to make contact with the patient to create a dangerous voltage drop across the body.
The operational control of the scanner is implemented with multiple safety mechanisms and features an independent monitoring of relevant metrics and has the ability to intervene in the signal generation. For example, the \ac{SU} is capable of disabling the transmission chain in the event of overheating. The same applies to the resonant tuning heating unit. An active imaging phase is only entered if no unexpected \ac{DF} feedback is observed.
Unintended high \ac{DF} levels due to incorrect inputs or component failures are prevented by the chosen operating point at the upper power limit of the amplifier. In addition, the resonant transmit chain is frequency specific and unexpected waveform changes or detuning of components in the transmit chain will reduce the power in the \ac{DFG}.
Finally, a human operator can use a hardware console to disable parts of the system at any point.

\subsection{Hardware Implementation}
The design concept for signal generation and reception, includes a symmetric approach to enable simultaneous signal generation and reception~\cite{sattel_setup_2015}, called send-receive approach (TxRx). This utilizes a pick-up node within the \ac{HCC} that is sensitive to the voltage signal induced by the particles magnetization response, while suppressing the excitation voltage.
The implementation of this approach requires an \ac{ICN} as one fundamental part. If built symmetrically with a similar serial resistance of the inductor to the \ac{DFG}, twice the power is required to match the field generated by a single resonance circuit, limiting the maximum \ac{DF} amplitude of the TxRx approach. 
In addition, the complexity of the send-receive chain increases, leading to a high susceptibility to disturbances generated within the \ac{HCC} (e.g. from connections and eddy currents). 
In our case, the receive signal acquired by this approach was inferior to the receive signal of a dedicated receive coil, which is able to suppress signal distortions~\cite{paysen_improved_2018}, justifying the utilization of the dedicated $x$-receive coil for this channel. 
A dedicated receive coil in the $z$-direction has not yet been developed due to the laborious design and intricate tuning process required for the cancellation coil. Due to the \ac{DF} saddle coil, the magnetic field profiles in $z$ have only a small homogeneous area and coupling to other channels becomes an issue.

The sampling trajectory originates from a superposition of the slowly varying selection field and the fast oscillating drive field. The arising \ac{FFP} movement follows a trajectory, which samples the \ac{FOV} and ideal \ac{MPI} systems try to generate homogeneous and orthogonal drive fields with a linear selection field. However, to achieve low field imperfections, large coils are required that are less energy efficient. Restrictions of design space, power supply and coil coupling~\cite{von_gladiss_influence_2018} within the \ac{DFG} of our scanner cause noticeable field imperfections as shown in \autoref{fig:fieldProfile}, that deviate from the ideal field. Especially towards the edges of the coils, field imperfections are severe and cause a deformation of the Lissajous type trajectory in all spatial dimensions. Our magnetic field measurements contain systematic uncertainties due to errors of coil sensor size and orientation. Consequently, they can only shear and rotate the resulting trajectory and the observed deformations can be attributed mainly to the imperfections of the magnetic fields.

\subsection{Conclusion and Outlook}

We have performed a comprehensive system characterization of a 3D human-sized \ac{MPI} scanner for real-time cerebral applications. The perfusion experiments conducted provide a proof-of-concept that the discrimination of brain hemispheres and different severities of stenosis are possible. 
This will allow us to better assess which clinical application scenarios are feasible in the future. 
For example, MPI trials for the ischemic stroke scenario in human volunteers may soon be possible using the \textit{Resotran} tracer. 
Tailored MPI tracers, such as  \textit{perimag} and \textit{synomag}, with future clinical approvals, promise to increase system performance and further expand the range of applications.

\section*{Acknowledgments}

We thank Florian Sevecke for technical support during the realization of numerous scanner components. We further thankfully acknowledge the financial support by the German Research Foundation (DFG, grantnumber KN 1108/7-1 and GR 5287/2-1). We also thank the developers of Makie.jl supporting us by answering technical questions when creating \autoref{fig:SMMultiPatch} and \autoref{fig:SMSinglePatch}.
Finally, we are grateful to Christian Findeklee for discussions and insights into resonant circuit decoupling.

\section*{Author Contributions}

F.T., F.F., F.M., T.K., and M.G. contributed to the system's conceptualization. F.T., F.F., and F.M. constructed the MPI components. N.H. and T.K. developed the software. F.T., F.F., F.M., N.H., M.B., M.M., and T.K.  contributed to experiment planning and execution. J.S. provided assistance with the reception system. T.K. and M.G. supervised the project. F.T., F.F., F.M., and T.K. contributed writing the paper with support of N.H., M.B., and M.M. All authors reviewed the manuscript.

\section*{Additional Information}

\noindent \textbf{Competing interests:} The authors declare no competing interests.

\noindent \textbf{Data availability:} Data sets generated during the current study are available from the corresponding author on reasonable request.

\bibliography{ref}

\end{document}

%% file: acronyms.tex
\usepackage[nolist,nohyperlinks]{acronym}

\begin{acronym}[ECU]

\acro{AWMPS}[AWMPS]{arbitrary waveform magnetic particle spectrometer}
\acro{ADC}[ADC]{analog-to-digital converter}
\acro{AUC}[AUC]{area under the curve}

\acro{CT}[CT]{computed tomography}

\acro{DAC}[DAC]{digital-to-analog converter}
\acro{DFG}[DFG]{drive-field generator}
\acro{DFCs}[DFCs]{drive-field coils}
\acro{DF}[DF]{drive-field}
\acro{DA}[DA]{differential amplifier}

\acro{ECD}[ECD]{equivalent circuit diagram}

\acro{FFL}[FFL]{field-free-line}
\acro{FFP}[FFP]{field-free-point}
\acro{FFR}[FFR]{field-free-region}
\acro{FFT}[FFT]{fast Fourier transform}
\acro{FOV}[FOV]{field-of-view}
\acro{FWHM}[FWHM]{full width at half maximum}

\acro{GUI}[GUI]{graphical user interface}

\acro{HCC}[HCC]{high current circuit}

\acro{ICN}[ICN]{inductive coupling network}
\acro{ISI}[ISI]{integrated signal intensity}
\acro{ICs}[ICs]{integrated circuits}
\acro{IA}[IA]{instrumentation amplifier}

\acro{ICU}[ICU]{intensive care unit} 
\acro{LFR}[LFR]{low-field-region}
\acro{LNA}[LNA]{low noise amplifier}

\acro{MPI}[MPI]{Magnetic Particle Imaging}
\acro{MRI}[MRI]{Magnetic Resonance Imaging}
\acro{MTT}[MTT]{mean-transit-time}

\acro{PSF}[PSF]{point spread function}
\acro{PNS}[PNS]{peripheral nerve stimulation}
\acro{PTT}[PTT]{pulmonary transit time}

\acro{Q}[Q-factor]{quality factor}

\acro{RF}[RF]{radio-frequency}
\acro{RPs}[RPs]{RedPitaya STEMlab 125-14}
\acro{RF-fields}[RF]{radio frequency fields}
\acro{rBV}[rBV]{relative blood-volume}
\acro{rBF}[rBF]{relative blood-flow}
\acro{rCBV}[rCBV]{relative cerebral-blood-volume}
\acro{rCBF}[rCBF]{relative cerebral-blood-flow}
\acro{RBCs}[RBCs]{red blood cells}

\acro{SNR}[SNR]{signal-to-noise ratio}
\acro{SU}[SU]{surveillance unit}
\acro{SAR}[SAR]{specific absorption rate}
\acro{SPIONs}[SPIONs]{superparamagnetic iron oxide nanoparticles}
\acro{MNPs}[MNPs]{magnetic nanoparticles}
\acro{SF}{selection field}
\acro{SFG}[SFG]{selection-field generator}

\acro{TF}[TF]{transfer function}
\acro{THD}[THD]{total harmonic distortion}
\acro{TTP}[TTP]{time-to-peak}

\acro{VOI}[VOI]{volume of interest}

\end{acronym}

%% file: math-definitions.tex
\usepackage{amsmath}
\usepackage{amsfonts}
\usepackage{amssymb}
\usepackage{nicefrac}
\usepackage{bbm} 
\usepackage{bm} 

\usepackage[all]{xy}

\newcommand{\IN}{\mathbb{N}}		
\newcommand{\IR}{\mathbb{R}}		
\newcommand{\IC}{\mathbb{C}}		

\renewcommand{\epsilon}{\varepsilon}				
\renewcommand{\theta}{\vartheta}                  	
\renewcommand{\phi}{\varphi}						

\newcommand{\set}[1]{\left\{ #1 \right\}}				
\newcommand{\norm}[1]{\left\lVert #1 \right\rVert}		

\newcommand{\signatur}[3]{#1:#2\,\rightarrow\,#3}     

\renewcommand{\ss}[1]{_\textup{\footnotesize #1}}

\usepackage[exponent-product = \cdot]{siunitx}
\DeclareSIUnit{\mup}{\text{$\mu_0$}}
\DeclareSIUnit{\mT}{\milli\tesla}
\DeclareSIUnit{\Tpm}{\tesla\per\meter}

\usepackage{amsthm}
\theoremstyle{plain} 

\theoremstyle{definition} 

\theoremstyle{remark} 

\usepackage{tikz}
\usepackage{tikz-3dplot}
\usepackage{pgfplots}
\usepackage{pgfplotstable}

\pgfplotsset{compat=1.8}

\usetikzlibrary{shapes,shapes.multipart,positioning}
\usepgfplotslibrary{units,statistics}

\pgfplotsset{
    unit markings=slash space,
    /pgfplots/xbar/.style={
    /pgf/bar shift={-0.5*(\numplotsofactualtype*\pgfplotbarwidth + (\numplotsofactualtype-1)*#1) + (.5+\plotnumofactualtype)*\pgfplotbarwidth + \plotnumofactualtype*#1},
    },
}
\pgfplotsset{select coords between index/.style 2 args={
    x filter/.code={
        \ifnum\coordindex<#1\fi
        \ifnum\coordindex>#2\fi
    }
}}

\usepackage{color}
\definecolor{ibilight}{RGB}{193,216,237}
\definecolor{ibidark}{RGB}{0,73,146}	
\definecolor{uke2}{RGB}{170,156,143} 	
\definecolor{uke3}{RGB}{87,87,86}		
\definecolor{ukesec1}{RGB}{255,223,0}	
\definecolor{ukesec2}{RGB}{239,123,5}	
\definecolor{ukesec3}{RGB}{104,195,205}	
\definecolor{ukesec4}{RGB}{138,189,36}	
\definecolor{ukesec5}{RGB}{178,34,41}	
\definecolor{tuhh}{RGB}{45,198,214}     
\definecolor{ibidarkBG}{RGB}{227,229,242}   
\definecolor{uke2BG}{RGB}{233,228,225} 	    
\definecolor{uke3BG}{RGB}{230,231,232}	    
\definecolor{ukesec1BG}{RGB}{255,243,190}   
\definecolor{ukesec2BG}{RGB}{254,232,212}   
\definecolor{ukesec3BG}{RGB}{222,241,241}   
\definecolor{ukesec4BG}{RGB}{233,243,222}   
\definecolor{ukesec5BG}{RGB}{244,230,225}   

\usepackage[color=ibilight]{todonotes}		